# Constitutive modeling of the tension-compression behavior of gradient structured materials

Jianfeng Zhao[a], Xiaochong Lu[a], Jinling Liu[a], Chen Bao[a], Guozheng Kang[a], Michael Zaiser[a,b], Xu Zhang [a*]

[a] *Applied Mechanics and Structure Safety Key Laboratory of Sichuan Province, School of Mechanics and Engineering, Southwest Jiaotong University, Chengdu 610031, China*

[b] *WW8-Materials Simulation, Department of Materials Science, Friedrich-Alexander-Universität Erlangen-Nürnberg, 90762 Fürth, Germany*

**Correspondent author: Tel: 86 (0)28 87601442, E-mail:* xzhang@swjtu.edu.cn

## Abstract

Gradient structured (GS) metals processed by severe plastic deformation techniques can be designed to achieve simultaneously high strength and high ductility. Significant kinematic hardening is key to their excellent strain hardening capacity which results in a favorable strength-ductility combination. Unfortunately, no constitutive model has been established to simulate and analyze the characteristic kinematic hardening behavior of GS metal to understand the relationship between their microstructure and macroscopic response. In this work, we developed a deformation-mechanism-based strain gradient plasticity model considering the plasticity heterogeneities from the grain to the sample scale. A back stress model, which accounts for the dependency of dislocation pile-ups on grain size, is established to describe the cyclic deformation properties of GS materials. The established model unified the geometrically necessary dislocations accommodating internal plasticity heterogeneities, the resulting back stress and reversible dislocations during reverse loading into a strain gradient plasticity framework, without introducing expedient independent material parameters. A finite element implementation of the model quantitatively predicts the uniaxial tensile and tensile-compressive responses of a GS copper bar as well as of a reference sample with homogeneous grain size. It is found that GS copper exhibits enhanced kinematic hardening which results mainly from fine grains in the GS layer and contributes to the considerable ductility of the GS material. The model allows to investigate the mechanical response and optimize the properties of materials with various types of spatially heterogeneous grain microstructures.

## 1. Introduction

Strength and ductility are among the most important indicators for the mechanical performance of structural materials, and combining high strength and high ductility in one material has been an important goal of materials research over many decades. Unfortunately, the two properties tend to be mutually exclusive. For example, coarse-grained (CG) metals usually exhibit satisfactory ductility, but low yield strength. When the grains in metals are refined to the submicron- or nano-scale, fewer dislocations pile up in front of grain boundaries due to the diminishing space. As a consequence, larger stress is required to assist the leading dislocation of a pile-up in overcoming the grain boundary barrier and the yield strength may increase significantly, in a manner that is well described by the Hall-Petch relation (Hall, 1951; Petch, 1953). On the other hand, dislocation behavior in such ultrafine-grained (UFG) or nano-grained materials is dominated by their emission and absorption at grain boundaries (Meyers et al., 2006). The diminishing importance of dislocation multiplication and storage in the grain interiors reduces the strain hardening capability of the material. As a result, metals with finer grains often suffer from reduced ductility, which may be of the order of a few percent plastic strain only. This poses serious restrictions to the use of nanostructured metals in industrial applications. Therefore, a major challenge is to improve the strength of a material while preserving a sufficient degree of ductility.

Inspired by natural materials, such as bone, bamboo, and shells, which often exhibit spatially graded microstructures, gradient structured (GS) metals have been synthesized and shown to exhibit high strength together with a considerable degree of ductility (Cheng et al., 2018; Fang et al., 2011; Lin et al., 2018; Long et al., 2019; Wei et al., 2014; Wu et al., 2014a). GS materials exhibit spatially heterogeneous grain size distributions, with grain sizes changing from the submicron or nano-scale in the near-surface region to the micrometer scale in the core of a sample. The unique microstructure of GS materials enables scientists to optimize them by controlling the grain size distribution or the volume fraction of the graded region. However, two issues need to be addressed before systematic approaches can be used to design GS materials for optimizing strength-ductility synergy: (1) One needs to reveal the physical mechanisms responsible for the high strength and considerable ductility of GS materials. (2) One needs to formulate quantitative models to predict the response of candidate GS microstructures under general loading conditions.

Obviously, the main contribution to the elevated yield strength of GS materials comes from regions with small grain sizes. In such regions, strength is increased according to the Hall-Petch relation, i.e., the nano/ultrafine grains in the graded structure enhance the yield strength of GS material through grain boundary strengthening. Another important factor affecting the deformation behavior of GS materials is the magnitude and spatial distribution of dislocation density in the as-manufactured material. Grain refinement by severe plastic deformation may introduce very high initial dislocation densities (Byer and Ramesh, 2013; Wang et al., 2019a; Zhou et al., 2017), a factor which further enhances strength but reduces strain hardening capability (Wang et al., 2019a). Recently, high dislocation density gradients were found in the gradient layer of GS materials treated by surface mechanical attrition treatment (SMAT) (Bahl et al., 2017; Kalsar and Suwas, 2018; Moering et al., 2016). In addition to the effects of grain size and initial dislocation density, synergetic strengthening mechanisms such as back stress and stress gradient effects have been discussed that may further enhance the yield stress (Moering et al., 2016; Wang et al., 2018; Wu et al., 2014b; Yang et al., 2016).

Despite the elevated yield stress, GS materials may exhibit significant strain hardening, which is key to their considerable ductility and thereby to the superior strength-ductility combination. Generally, under tensile loading, a freestanding gradient layer loses stability and fails by necking at a relatively small strain due to the lack of strain hardening ability that is inherent in material with small grain size and high dislocation density. However, during the deformation of an integrated GS structure, the necking instability of the surface gradient layer is constrained by the CG core so that the GS material can undergo further deformation (Wu et al., 2014a; Yuan et al., 2019). Besides, strain gradients accommodated by geometrically necessary dislocations (GNDs) and the emergence of multi-axial stress states promote dislocation multiplication and storage, and thus further enhance strain hardening. Another factor that may contribute considerably to strain hardening is the kinematic hardening due to dislocation induced back stresses. Dislocations pile up in front of grain boundaries to accommodate the deformation heterogeneities during the deformation of GS materials (Wu and Zhu, 2017; Zhu and Wu, 2019). The piled-up dislocations produce high back stresses to compensate the strength mismatch between large soft grains and the surrounding hard grains (Wu et al., 2015; Wu and Zhu, 2017); at the same time, they contribute to forest hardening. Thus, the overall strength of the heterogeneous material increases. Generally, back stress hardening is also observed in homogeneously-grained metals, but more grain boundaries are introduced and larger strain heterogeneity arises in GS materials due to the plastic deformation incompatibilities that require deformation accommodation not only between adjacent grains but also on the larger scale of the grain size gradient, leading to enhanced back stress levels. Several experiments have demonstrated the extraordinary kinematic hardening in GS materials. For example, Yang et al. (Yang et al., 2016) determined the back stress of a GS IF-steel sheet by performing an unloading-reloading procedure; they found that the back stress accounts for about 35% of the total stress. Liu et al. (Liu et al., 2018) also observed a more pronounced Bauschinger effect in GS copper samples than in their CG counterparts.

Complementary to experimental investigation, theoretical modeling is capable of providing quantitative expressions that formulate relationships between the microstructure and macroscopic mechanical response of GS materials (Li and Soh, 2012; Li et al., 2017; Lu et al., 2019; Zhao et al., 2019a; Zhao et al., 2019b; Zhu and Lu, 2012), allow to analyze their response under complex loading conditions, and help in designing their microstructures. For instance, Zhao et al. (Zhao et al., 2019a) developed a dislocation-density-based model considering the deformation mechanisms of grains with different sizes and successfully predicted the relations among the strength, ductility, and microstructures of GS materials. Lu et al. (Lu et al., 2019) obtained similar results using a crystal plasticity method coupling with a homogenization scheme. However, the above models are usually constructed exclusively for predicting the uniaxial tensile response of GS materials, while the extraordinary back stress hardening revealed by experiments (Liu et al., 2018; Wu et al., 2015; Yang et al., 2016) has always been neglected, and thus the constitutive model for describing the cyclic deformation is lacking.

A lot of sophisticated models were developed to describe the kinematic hardening and the resulting Bauschinger effects characterizing the change of yield stress when loading is reversed (Armstrong and Frederick, 1966; Chaboche, 2008; Kang and Kan). Although these models have achieved great success in describing the mechanical responses of materials under complex loading conditions, the parameters used were always phenomenological and needed to be determined by conducting heavy mechanical tests. Taken the GS materials as an example, since the grain sizes inside spans over three to four orders of magnitude, back stress parameters for homogeneous materials with different grain sizes have to be obtained by performing extensive experiments, which is not only time consuming but also discommodious for application. Furthermore, in order to capture the non-linearity of stress-strain curves during unloading or the transient in strain hardening rate when loading is reversed, an extra term namely reversible dislocation density has been introduced (Castelluccio and McDowell, 2017; Kitayama et al., 2013; Rauch et al., 2007; Wen et al., 2015; Zecevic and Knezevic, 2015). Although the reversible dislocations have explicit physical meaning, i.e. the "erase" of the trapped dislocations during the prestrain facilitated by back stress when the loading is reversed (Mompiou et al., 2012; Rauch et al., 2007), the evolution law of their density was always constructed independently and the related back stress evolution law was also established empirically (Wen et al., 2015; Zecevic and Knezevic, 2015). As a consequence, more phenomenological parameters were introduced.

The objective of this work, therefore, is to develop a physically-based model with as few expedient parameters as possible to reveal the strain hardening mechanisms and predict the cyclic plasticity response of GS materials. To achieve this goal, in Section 2, a deformation-mechanism-based model considering the plasticity heterogeneities from the grain to the sample scale is established. Then the developed model is implemented into a finite element framework to simulate the tension-compression responses of GS copper as well as CG reference samples in Section 3. A comparison between simulation data and experimental results and corresponding data analysis are presented in Section 4. The study ends with some conclusions.

## 2. A constitutive model for GS materials

The distinctive feature of GS materials as compared to CG ones is the spatially graded grain size. The deformation incompatibilities between grains and different parts of GS materials are accommodated by GNDs, which, in turn, contribute to isotropic hardening. To model the deformation mechanisms controlling the strain hardening of GS materials as well as their dependence on grain size effects, a conventional mechanism-based strain gradient plasticity (CMSG) model developed by Huang et al. (Huang et al., 2004) was used as starting point. This dislocation-based model was modified to account for the internal deformation heterogeneities resulting from the heterogeneous microstructure. These heterogeneities are described in terms of piled-up GNDs and associated back stresses, to obtain a model of kinematic hardening that allows describing the cyclic deformation of GS materials faithfully.

### 2.1. The framework of the CMSG model

In our discussion of the CMSG model, we follow Huang et al. (Huang et al., 2004). For an elasto-plastic solid at small deformation, the strain rate can be decomposed into elastic and plastic parts,

$$\dot{\boldsymbol{\varepsilon}} = \dot{\boldsymbol{\varepsilon}}^{\mathrm{e}} + \dot{\boldsymbol{\varepsilon}}^{\mathrm{p}} . \tag{1}$$

For isotropic material, Hooke's law describes the relationship between elastic strain rate and stress rate as

$$\dot{\boldsymbol{\varepsilon}}^{\mathrm{e}} = \frac{1}{2\mu}\dot{\boldsymbol{\sigma}}' + \frac{1}{9K}\mathrm{tr}\left(\dot{\boldsymbol{\sigma}}\right)\boldsymbol{I} , \tag{2}$$

where $\dot{\boldsymbol{\sigma}}' = \dot{\boldsymbol{\sigma}} - \mathrm{tr}\left(\dot{\boldsymbol{\sigma}}\right)\boldsymbol{I}/3 = 2\mu\left(\dot{\boldsymbol{\varepsilon}}' - \dot{\boldsymbol{\varepsilon}}^{\mathrm{p}}\right)$ is the deviatoric stress rate, $\dot{\boldsymbol{\sigma}}$ is the stress rate, $\boldsymbol{I}$ is the unit tensor with components $I_{ij} = \delta_{ij}$, $\dot{\boldsymbol{\varepsilon}}' = \dot{\boldsymbol{\varepsilon}} - \mathrm{tr}\left(\dot{\boldsymbol{\varepsilon}}\right)\boldsymbol{I}/3$ is the deviatoric strain rate, $\mu$ and $K$ are shear modulus and volume modulus, respectively.

The plastic strain rate is proportional to the deviatoric stress rate $\dot{\boldsymbol{\sigma}}'$ according to the $J_2$-flow theory,

$$\dot{\boldsymbol{\varepsilon}}^{\mathrm{p}} = \frac{3\boldsymbol{\sigma}'}{2\bar{\sigma}}\dot{p} , \tag{3}$$

where $\dot{p} = \sqrt{2\dot{\boldsymbol{\varepsilon}}^{\mathrm{p}} : \dot{\boldsymbol{\varepsilon}}^{\mathrm{p}} / 3}$ is the effective plastic strain rate and $\bar{\sigma} = \sqrt{3\boldsymbol{\sigma}' : \boldsymbol{\sigma}' / 2}$ is the effective stress. A visco-plastic formula relates $\dot{p}$ to the effective stress,

$$\dot{p} = \dot{\varepsilon}_0 \left(\frac{\bar{\sigma}}{\sigma_{\mathrm{f}}}\right)^m , \tag{4}$$

where $\dot{\varepsilon}_0$ is a reference strain rate, which was modified to equal the effective deviatoric strain rate $\dot{\bar{\varepsilon}} = \sqrt{2\dot{\boldsymbol{\varepsilon}}' : \dot{\boldsymbol{\varepsilon}}' / 3}$ by Kok et al. (Kok et al., 2002) to eliminate the explicit time dependence, and $\dot{\varepsilon}' = \dot{\varepsilon} - \mathrm{tr}(\dot{\varepsilon})\boldsymbol{I}/3$ is the deviatoric strain rate. If $m$ takes a sufficiently large value (e.g., larger than 20), the formulation approaches an elasto-plastic one. $\sigma_{\mathrm{f}}$ is the flow stress of the material under uniaxial tension, which is material dependent and governed by the deformation mechanisms.

When the plastic deformation of metals is controlled by dislocation motion, the flow stress can be described by the Taylor hardening law (Taylor, 1934), which relates the critical resolved shear stress $\tau$ on the active slip system(s) to the dislocation density $\rho$ via

$$\tau = \alpha\mu b\sqrt{\rho}, \tag{5}$$

where $\alpha$ is an empirical coefficient usually taken as 0.3, $b$ denotes the magnitude of the Burgers vector, which is 0.256 nm for copper. The macroscopic flow stress can be obtained from the shear stress in the active slip systems by

$$\sigma_{\mathrm{f}} = M\alpha\mu b\sqrt{\rho}. \tag{6}$$

where the averaged orientation factor $M$ (Taylor factor) equals 3.06 for FCC metals. In order to incorporate the contribution of strain gradients to strain hardening into the above model, the dislocation density $\rho$ is decomposed into two parts: the density of statistically stored dislocations (SSDs) and the density of GNDs, i.e.,

$$\rho = \rho_{\mathrm{SSD}} + \rho_{\mathrm{GND}}. \tag{7}$$

where SSDs accumulate by randomly trapping each other, and GNDs accommodate the non-uniform plastic deformation. So, Eq. (6) is re-written as

$$\sigma_{\mathrm{f}} = M\alpha\mu b\sqrt{\rho_{\mathrm{SSD}} + \rho_{\mathrm{GND}}}. \tag{8}$$

It is hypothesized that $\rho_{\mathrm{SSD}}$ can be derived from the uniaxial tension response of materials which eliminates the influence of GNDs, while $\rho_{\mathrm{GND}}$ is related to the strain gradient as

$$\rho_{\mathrm{GND}} = \bar{r}\frac{\eta^{\mathrm{p}}}{b}, \tag{9}$$

where $\bar{r}$ is the Nye factor. The effective plastic strain gradient $\eta^{\mathrm{p}}$ is defined as

$$\eta^{\mathrm{p}} = \sqrt{\frac{1}{4}\eta^{\mathrm{p}}_{ijk}\eta^{\mathrm{p}}_{ijk}}, \tag{10}$$

where $\eta_{ijk}^{\mathrm{p}} = \varepsilon_{ik,j}^{\mathrm{p}} + \varepsilon_{jk,i}^{\mathrm{p}} - \varepsilon_{ij,k}^{\mathrm{p}}$ [23].

The CMSG model integrates the GNDs-based Taylor hardening model into the conventional $J_2$-flow theory to describe the strain gradient effect. Furthermore, various dislocation evolution laws can be incorporated to achieve a physically-based description of the underlying deformation mechanisms. However, conventional strain gradient plasticity theories, including the CMSG model described here, were mostly used to investigate the behavior of materials with homogeneous microstructure under non-uniform deformation, such as wire torsion (Fleck et al., 1994; Liu et al., 2012), foil bending (Stölken and Evans, 1998) and nanoindentation (Nix and Gao, 1998), but much less to investigate internal deformation inhomogeneity resulting from heterogeneous microstructure. Furthermore, in the original CMSG model, only isotropic hardening controlling the extension of the elastic domain is considered. In the following, we extend the CMSG model to a more general form by considering, in addition to GNDs accommodating sample-scale strain inhomogeneities, also the GNDs that accommodate strain inhomogeneity between neighboring grains. These GNDs, which take the form of pile-ups at grain boundaries, contribute both to forest hardening by acting as obstacles to dislocations on other slip systems, and to kinematic hardening by producing long-range back stresses. The latter aspect is of particular importance in describing cyclic deformation behavior.

### 2.2. CMSG model considering internal deformation heterogeneities in GS materials

To model the grain size effects governing the initial yielding and extraordinary Bauschinger effect in GS materials, we need a framework that accounts for both kinematic and isotropic hardening contributions and captures the influence of grain size. To this end, Eq. (8) describing the flow stress and Eq. (3) dominating the plastic deformation need to be modified.

#### *2.2.1. Initial yield stress*

The grain size effect can be incorporated into Eq. (8) in terms of the well-known Hall-Petch relation as

$$\sigma_{\mathrm{f}} = \sigma_0 + k_{\mathrm{HP}} d^{-1/2} + M\alpha\mu b\sqrt{\rho_{\mathrm{SSDs}} + \rho_{\mathrm{GNDs}}} \,. \tag{11}$$

The first two terms on the right-hand side represent the Hall-Petch formulation $\sigma_{\mathrm{Y}} = \sigma_0 + k_{\mathrm{HP}} d^{-1/2}$, where $\sigma_0$ is the lattice friction stress, $k_{\mathrm{HP}}$ is the Hall-Petch constant determined by experimental results and $d$ denotes the grain size.

### *2.2.1. Kinematic hardening*

To describe the cyclic deformation of the materials, the flow rule is modified to consider the influence of back stress on plastic deformation, i.e., we modify Eq. (3) to

$$\dot{\boldsymbol{\varepsilon}}^{\mathrm{p}} = \frac{3\dot{p}}{2\sigma_{\mathrm{e}}}\left(\boldsymbol{\sigma}' - \boldsymbol{\sigma}^{\mathrm{b}}\right), \tag{12}$$

where $\boldsymbol{\sigma}^{\mathrm{b}}$ denotes the back stress tensor. Accordingly, the definition of effective stress changes to

$$\bar{\sigma} = \sqrt{3\left(\boldsymbol{\sigma}' - \boldsymbol{\sigma}^{\mathrm{b}}\right):\left(\boldsymbol{\sigma}' - \boldsymbol{\sigma}^{\mathrm{b}}\right)/2}\,. \tag{13}$$

Physically, the back stress results mainly from the pile-up of dislocations. As a single-ended pile-up schematically shown in Fig. 1, dislocations pile up when they move towards obstacles such as grain boundaries, precipitates, and twin boundaries. Here we focus on piled-up dislocations at grain boundaries that accommodate slip discontinuities between adjacent grains. It should be noted that although other dislocation structures may accommodate strain heterogeneities between adjacent grains (Ashby, 1970), only pile-up configurations are considered in this work for clarity and simplicity. Under loading, the piled-up dislocations produce back stresses that impede the motion of forthcoming dislocations. As a consequence, higher external stress is required to activate further plastic deformation. When the loading reverses, the slip directions of dislocations also reverse, hence during reverse loading the pile-up stresses facilitate the reverse motion of dislocations. Therefore, lower external stress is needed to drive dislocation glide. Apart from this, other mechanisms that may contribute to the non-linearity of unloading and the apparent Bauschinger effect were also proposed (Zecevic and Knezevic, 2015), such as the annihilation of trapped dislocations during unloading and the intra-granular back stress caused by the incompatible deformation of dislocation cell-wall structures inside grains.

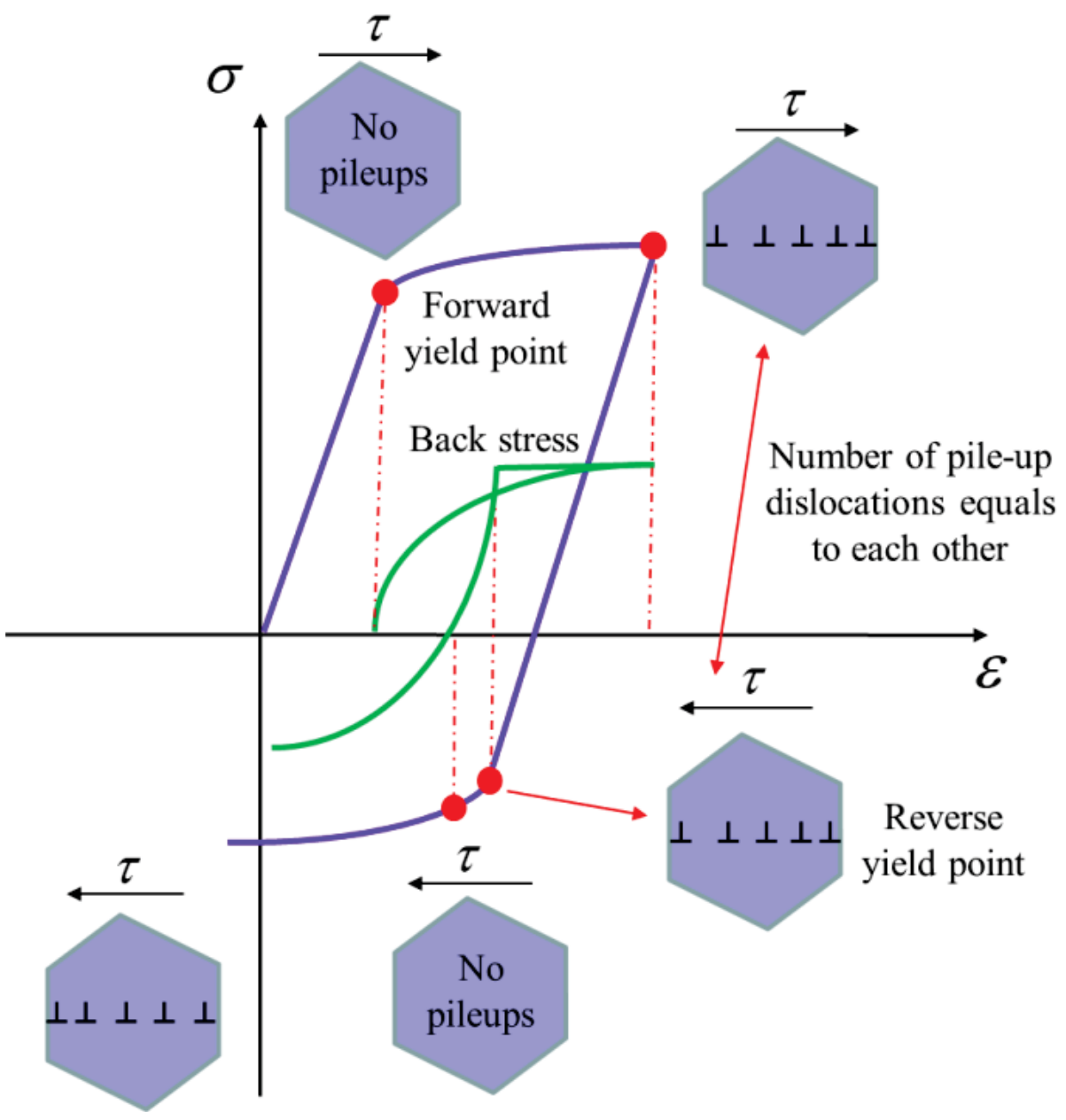

**Fig. 1.** Schematic representation of cyclic stress-strain curves, back stress evolution, and behavior of piled-up dislocations during cyclic deformation.

Based on this physical picture, and assuming the dislocations in the pile-up to be of edge type exclusively, we use a tensorial generalization of scalar back stress models and express the back stress in terms of the number of pile-up GNDs and grain size as,

$$\boldsymbol{\sigma}^{\mathrm{b}} = \frac{M\mu b}{\pi(1-\nu)d}\boldsymbol{N}, \tag{14}$$

where $\nu$ is Poisson's ratio, $n = \sqrt{\boldsymbol{N}:\boldsymbol{N}}$ is the characteristic number of dislocations in a pile-up, and the unit tensor $\boldsymbol{N}/n$ defines the direction of the associated back stress tensor. To provide an equation of evolution for $\boldsymbol{N}$ we generalize the model of Sinclair et al. (Sinclair et al., 2006) to tensorial form,

$$\dot{\boldsymbol{N}} = N_{\Delta}\left(\frac{2}{3}\dot{\boldsymbol{\varepsilon}}^{\mathrm{p}} - \frac{\boldsymbol{N}}{N^{*}}\dot{p}\right), \tag{15}$$

where $N_{\Delta}$ is the initial growth rate of $\boldsymbol{N}$, and $N^{*}$ is the maximum number of dislocations in the pile-up. $N_{\Delta}$ and $N^{*}$ are here assumed to be scalar (direction independent) quantities. From Eq. (14) and (15), $N^{*}$ controls the saturated value of back stress, while $N^{*}$ and $N_{\Delta}$ control the characteristic plastic strain required for back stress changes. In the following, the expressions for $N^{*}$ and $N_{\Delta}$ are determined using the physical picture of a single-ended pile-up.

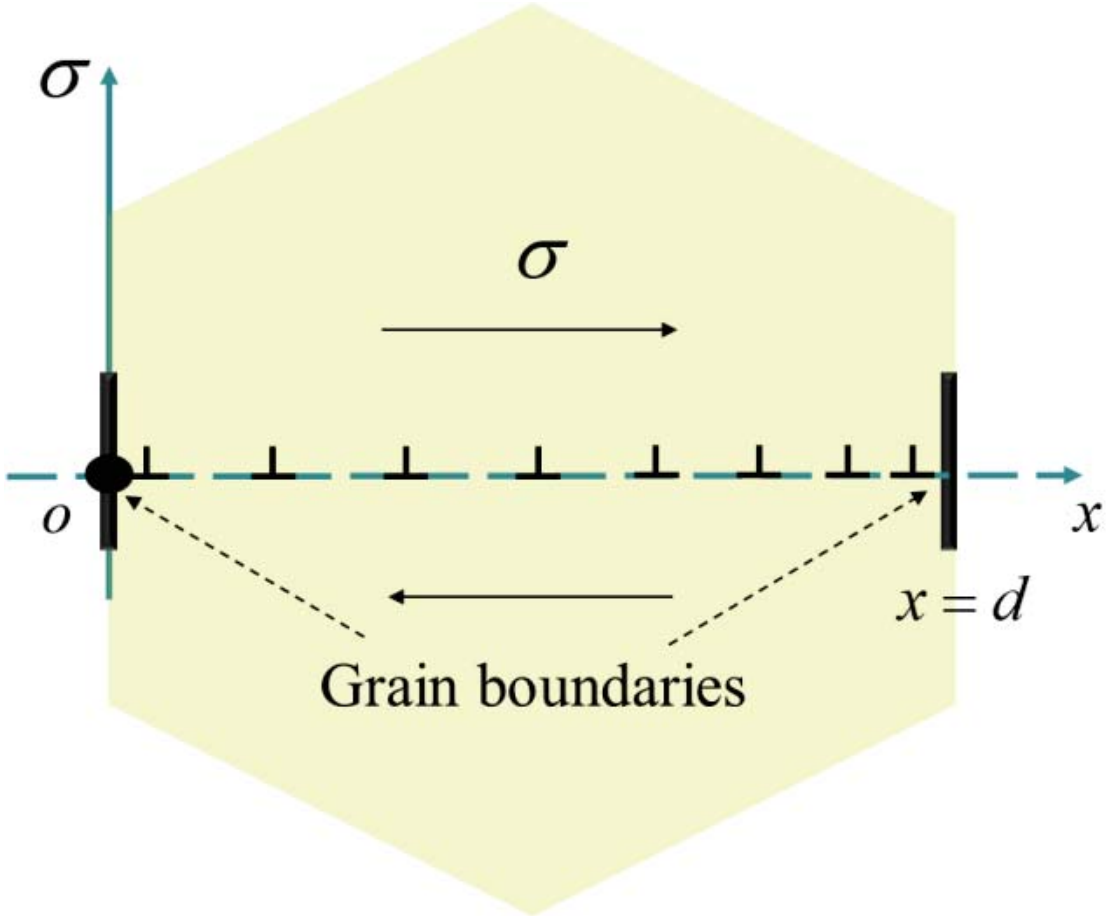

**Fig. 2.** Schematic of the single-ended pile-up under applied stress.

Based on a single-ended pile-up picture as shown in Fig. 2, the relation among the number of dislocations *N* in a pile-up, the applied stress $\sigma$ and the grain size *d* given by Li et al. (Li and Chou, 1970) is rewritten as

$$N = \frac{\pi(1-\nu)(\sigma-\sigma_0)d}{\mu b}, \tag{16}$$

if the lattice friction stress is considered. When the stress acting on the tip of the pile-up reaches a critical stress level $\sigma_c$, the grain boundary barrier is considered to be overcome, and *N* reaches the maximum value $N^*$. Therefore, $N^*$ can be given as,

$$N^* = \frac{\pi(1-\nu)k_{HP}}{\mu b}d^{\frac{1}{2}}. \tag{17}$$

We now assume that several sources are activated and form pile-ups under the applied stress $\sigma$. If the locations of these sources are statistically independent, then it can be shown that the average interaction between the pile-ups is zero (Zaiser, 2013), and the above stated relations for a single pile-up remain unchanged. If the average distance between the ensuing slip lines is $\lambda$, then the total plastic strain produced in a grain of size *d* by dislocation glide during pile-up formation can be expressed as

$$\varepsilon_{pp} = \frac{b}{\lambda d}\int_0^d f(x)x\mathrm{d}x, \tag{18}$$

where $f(x)=2(\sigma-\sigma_0)(1-\nu)\sqrt{x/(l-x)}/(\mu b)$ is the dislocation density at position *x*. $N_\Delta$ can be approximated as

$$N_\Delta=\frac{N}{\varepsilon_{pp}}. \tag{19}$$

After integral of Eq. (18) and substituting Eqs. (16) and (18) into Eq. (19) give

$$N_\Delta=\frac{4\lambda}{3b}. \tag{20}$$

This result is similar to that adopted by Sinclair et al. (Sinclair et al., 2006), $\lambda/b$, and here the detailed pile-up configuration is considered. Substituting Eqs. (17) and (20) into Eq. (15), then combining Eqs. (14) and (15) we have

$$\dot{\boldsymbol{\sigma}}^{b}=\frac{4M\mu\lambda}{3\pi(1-\nu)d}\left(\frac{2}{3}\dot{\boldsymbol{\varepsilon}}^{p}-\frac{\mu b\boldsymbol{N}}{\pi(1-\nu)k_{HP}\sqrt{d}}\dot{p}\right)=\frac{8M\mu\lambda}{9\pi(1-\nu)d}\dot{\boldsymbol{\varepsilon}}^{p}-\frac{4\mu\lambda}{3\pi(1-\nu)k_{HP}\sqrt{d}}\boldsymbol{\sigma}^{b}\dot{p}, \tag{21}$$

Eq. (21) is similar to the widely used Armstrong-Frederick (A-F) kinematic hardening law (Armstrong and Frederick, 1966)

$$\dot{\boldsymbol{\sigma}}^{b}=C\dot{\boldsymbol{\varepsilon}}^{p}-\gamma\boldsymbol{\sigma}^{b}\dot{p}, \tag{22}$$

where the phenomenological parameters *C* and $\gamma$ can now be interpreted in physical terms via

$$C=\frac{8M\mu\lambda}{9\pi(1-\nu)d},\ \gamma=\frac{4\mu\lambda}{3\pi(1-\nu)k_{HP}\sqrt{d}}, \tag{23}$$

by which the dependence of back stress on grain size is established.

### *2.2.3. Evolution of* $\rho_{GNDs}$ *and* $\rho_{SSDs}$

As mentioned in Section 2.2.2, dislocations piled up near grain boundaries influence the mechanical response of GS materials in two different ways. On the one hand, they produce long-range back stresses that obstruct dislocation movement and cause kinematic hardening; on the other hand, they act as forest dislocations interacting with moving dislocations to cause further strengthening. Conventional strain gradient plasticity models usually focused on the strain hardening coming from GNDs required for accommodating strain gradients on the sample scale. In the present work, the piled-up dislocations accommodating slip discontinuities among grains are also considered part of the GND density. The number of GNDs in a pile-up is $n = \sqrt{\boldsymbol{N} : \boldsymbol{N}}$, and the number of pile-ups is $d/\lambda$, leading to a modification of Eq. (9):

$$\rho_{\text{GNDs}} = \bar{r}\frac{\eta^{\text{p}}}{b} + \frac{n}{\lambda d}, \tag{24}$$

the second term, which describes the density evolution of trapped dislocations near grain boundaries (here refer to pile-ups), has another benefit. Combining the second term in Eq. (24) with Eq. (15), it can be analyzed that during loading, dislocations are trapped to accommodate the deformation heterogeneities among grains, and thus, their density increases gradually; when the loading reverses, these pile-ups are released facilitated by back stress, and thus the corresponding dislocation density decreases to zero, indicating a dislocation annihilation process. This idea is similar to that adopted by Castelluccio et al. (Castelluccio and McDowell, 2017), where the annihilation of double-ended pileups of the opposite sign during unloading is utilized to describe the dislocation reversible process upon reverse loading. Therefore, physically, the second term in Eq. (24) can be employed to describe the evolution of reversible dislocation population, which has been usually introduced as an independent term by other researchers (Castelluccio and McDowell, 2017; Kitayama et al., 2013; Rauch et al., 2007; Wen et al., 2015; Zecevic and Knezevic, 2015). However, the models established here unifies the pile-ups accommodating the internal deformation heterogeneities among grains, the resulting back stress, and the reversible dislocations using correlative laws without inducing independent evolution equations and ad hoc parameters.

For the evolution of $\rho_{\text{SSDs}}$, we employ a modified KME model (Li and Soh, 2012) which considers the influence of grain boundaries on the multiplication and annihilation of dislocations,

$$\frac{\partial \rho_{\text{SSDs}}}{\partial p} = M\left[\frac{k_{\text{mfp}}^{\text{g}}}{bd} + \frac{k_{\text{mfp}}^{\text{dis}}}{b}\sqrt{\rho_{\text{SSDs}} + \rho_{\text{GNDs}}} - k_{\text{ann}}\left(\frac{\dot{p}}{\dot{\varepsilon}_{\text{ref}}}\right)^{-\frac{1}{n_0}}\rho_{\text{SSDs}} - \left(\frac{d_{\text{ref}}}{d}\right)^2\rho_{\text{SSDs}}\right], \tag{25}$$

where $k_{\text{mfp}}^{\text{g}}$ and $k_{\text{mfp}}^{\text{dis}}$ are proportionality factors, $k_{\text{ann}}$ is a material constant, $\dot{\varepsilon}_{\text{ref}}$ is a reference strain rate, and $n_0$ is related to temperature. $d_{\text{ref}}$ is a referenced grain size. The second and third terms in Eq. (25) constitute the original KME model, but the influence of GNDs on the mean free path of dislocations is incorporated in the second term. The first term represents the contribution of grain size $d$ to the mean free path of dislocations; the last term considers the enhanced annihilation of dislocations at grain boundaries as the grain size decreases. Thus $d_{\text{ref}}$ can be regarded as a reference grain size characterizing the annihilation of dislocations at grain boundaries.

In summary, we construct a dislocation-density-based model for describing the cyclic deformation behavior of GS materials in this Section. A deformation-mechanism-based model unifies the GNDs accommodating internal plasticity heterogeneities, the resulting back stress and reversible dislocation density is established, their dependence on grain size is constructed physically, and the influence of grain size on the SSDs' density is also considered. In the following, this model is applied to model the tension-compression response of GS copper.

## 3. Finite element model

The GS sample studied in this work is a SMAT copper bar with a diameter of 3 mm and a gauge length of 15 mm (Liu et al., 2018). The thickness of the GS layer is about 400 μm, with a gradient distributed grain size changing from 300 nm in the surface to 78.8 μm in the core along the radial direction, as shown in Fig. 3(a). The GS materials were usually treated as multi-layer composites with each layer being homogeneous, then the rule of mixture method was employed to obtain the overall mechanical response of GS materials (Jin et al., 2018; Li and Soh, 2012; Li et al., 2017). In this work, instead of the rule of mixture method, the finite element method is adopted to model GS materials since the constraint between layers can be better handled, and the strain gradient can be easily calculated. When conducting a finite element simulation, the finite element model needs to be geometrically consistent with the experimental one, and appropriate boundary conditions should also be applied according to the loading condition. Due to the symmetry of the bar, a 2D axisymmetric plane model is employed to model the 3D cylinder, as shown in Fig. 3(b). The depth of the model is 1.5 mm (exactly the radius of the specimen), the width is set as 400 μm. To represent the gradient distributed grain size, each integration point is endowed with a specific grain size based on experimental results (Liu et al., 2018). Obviously, the more refined the mesh is, the more accurate the grain size distribution can be reflected. Here, the finite element model is meshed with 6000 8-node symmetrical (CAX8R) elements. Axisymmetric boundary conditions are applied on the left and bottom sides, strain-controlled uniaxial tension (tension-compression for cyclic loading) at a rate of $5 \times 10^{-4}\ \text{s}^{-1}$ is applied on the upper side (Liu et al., 2018).

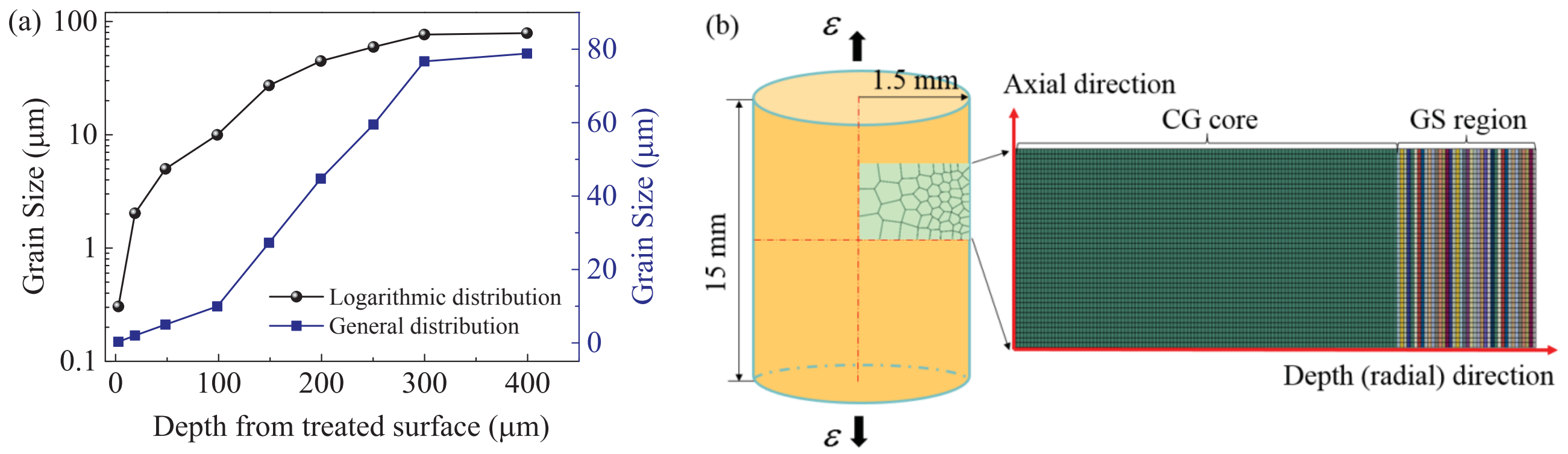

**Fig. 3.** (a) The grain size distribution along the depth direction. (b) The gradient structure is treated as a multi-layer composite, and a finite element model is constructed to mimic the experimental specimen.

With the finite element model constructed here and the grain-size-dependent model established in Section 2, the relationship between gradient microstructure and the macroscopic tensile response of GS material can be obtained via implementing the model in finite element simulation software ABAQUS (ABAQUS, 2014) utilizing the user material subroutine (UMAT). The numerical procedure for the finite element implementation of the developed model can be found in Appendix A.

## 4. Results and discussion

### 4.1. Validation of the proposed model

To model the overall mechanical response of GS copper, the dislocation-based model established in Section 2 is firstly validated by modeling homogeneously-grained coppers with different grain sizes. Uniaxial tensile responses of homogeneously-grained coppers with grain sizes of 500 nm, 25 μm, and 78.8 μm were modeled, as shown in Fig. 4(a). The corresponding experimental curves of true stress versus true strain are also shown for comparison (Liu et al., 2018). The uniaxial tension-compression response of the CG copper with a grain size of 78.8 μm was also simulated. The parameters used in the constitutive model are shown in Table. 1. Some parameters can be extracted from literature, such as the lattice friction stress $\sigma_0$ and the magnitude of the Burgers vector *b*. Other parameters, such as $k_{mfp}^{dis}$ controlling the dislocation multiplication and $k_{ann}$ related to the dislocation annihilation, can also be determined exclusively by simultaneous fitting to experimental results of homogeneously-grained copper with different grain sizes. The full parameterization process is detailed in Appendix B.

**Table 1.** Materials parameters for the constitutive model in Section 2.

| Parameter | Symbol | Value |
|---|---|---|
| Shear modulus (GPa) | $\mu$ | 42.1 |
| Lattice friction stress (MPa) | $\sigma_0$ | 25.5 |
| Hall-Petch constant 1 (MPa · μm$^{1/2}$) | $k_{HP}$ | 45 |
| Rate sensitively exponent | $m$ | 20 |
| Geometric factor | $k_{mfp}^{g}$ | 0.1 |
| Proportionality factor | $k_{mfp}^{dis}$ | 0.027 |
| Dynamic recovery constant 1 | $k_{ann}$ | 2.5 |
| Dynamic recovery constant 2 | $n_0$ | 21.25 |
| Reference strain rate ($s^{-1}$) | $\dot{\varepsilon}_{ref}$ | 1.0 |
| Reference grain size (μm) | $d_{ref}$ | 3.0 |
| Nye-factor | $\bar{r}$ | 1.9 |
| Distance between slip lines (μm) | $\lambda$ | 0.2 |

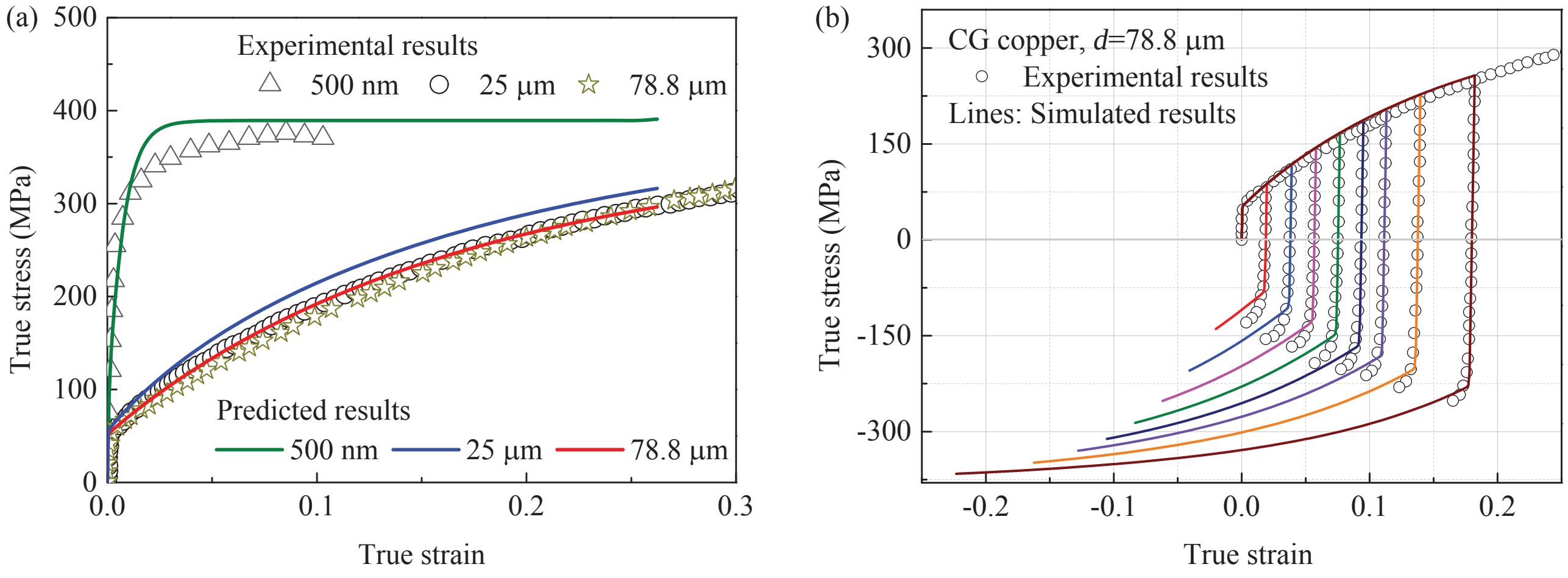

**Fig. 4.** (a) Simulated curves of uniaxial tensile true stress versus true strain of homogeneously-grained coppers and (b) tension-compression responses of CG copper. Experimental results (Fang et al., 2011; Liu et al., 2018; Lu et al., 2009) are also shown for comparison.

Fig. 4(a) shows that the simulated results using the established model agree well with the experimental results for copper with different grain sizes. The 0.2% offset yield stress ($\sigma_{0.2}$) increases from about 56 MPa to 196 MPa with a decrease of grain size from 78.8 μm to 500 nm. The strain hardening rate decreases with the decrease of grain size, which is consistent with the deformation mechanisms mentioned in Section 2, where larger grains possess more space for dislocation multiplication and storage. Moreover, the flow stresses for coppers with grain sizes of 25 μm and 78.8 μm display little difference with each other, indicating that when grain size lies in the CG region, the grain size has only limited influence on the tensile mechanical response. Fig. 4(b) shows the simulated tension-compression curves of CG copper and their comparison with experimental results. The simulated tension-compression responses at different strain levels are in good agreement with experimental results. Specifically, the inverse yield points characterizing the Bauschinger effect are well predicted for different pre-strains. Furthermore, it can be indicated that the kinematic hardening in CG copper is not strong. For example, the back stress at a true strain of 2% is about 3.8 MPa. When the true strain further increases to 18.2%, the back stress only increases to about 14.3 MPa, which is less than 10% of the overall flow stress. The good agreement between simulated results and experimental data, for both uniaxial tension and tension-compression, confirms the suitability of the proposed model for modeling copper with a wide range of grain sizes.

### 4.2. Tension-compression behavior of GS copper

#### 4.2.1. Initial yielding

As stated in the Introduction, during the preparation process of GS materials by severe plastic deformation, such as SMAT technique, large amounts of dislocations are introduced accompanied by the refinement of grains. Therefore, when the yield stress of GS materials is assessed, the influence of the high initial dislocation density has to be considered, and the yield stress is therefore

$$\sigma_{\mathrm{Y}} = \sigma_0 + k_{\mathrm{HP}} d^{-1/2} + M\alpha\mu b\sqrt{\rho_0}\ . \tag{26}$$

According to the experimental results (Bahl et al., 2017; Kalsar and Suwas, 2018; Moering et al., 2016), the initial dislocation density $\rho_0$ displays a gradient opposite to the grain size gradient: it decreases from the treated surface to the bulk. During mechanical testing, the initial dislocation density gradient influences not only the initial yielding but also the subsequent strain hardening behavior of the material. However, since spatially resolved experimental measurements of dislocation density are usually lacking, previous models often used piecewise constant initial dislocation densities, with different but spatially homogeneous density values in the gradient region and the CG core (Li and Soh, 2012; Zhao et al., 2019a). In the present work, we compare modeling results assuming spatially graded initial dislocation density with those that assume a piecewise constant initial dislocation density, as shown in Fig. 5(a). For the continuously graded case, dislocation density decreases from $8\times10^{15}/\mathrm{m}^2$ to $4\times10^{12}/\mathrm{m}^2$ from the surface to the core, where $10^{15}/\mathrm{m}^2$ is the magnitude of dislocation density in copper processed by severe plastic deformation (Gubicza et al., 2005; Ungár et al., 2011), and $4\times10^{12}/\mathrm{m}^2$ is the same value as used for CG copper in Section 4.1. For the piecewise constant case, a dislocation density of $1.5\times10^{15}/\mathrm{m}^2$ is adopted in the gradient layer to achieve a good agreement with experimental results.

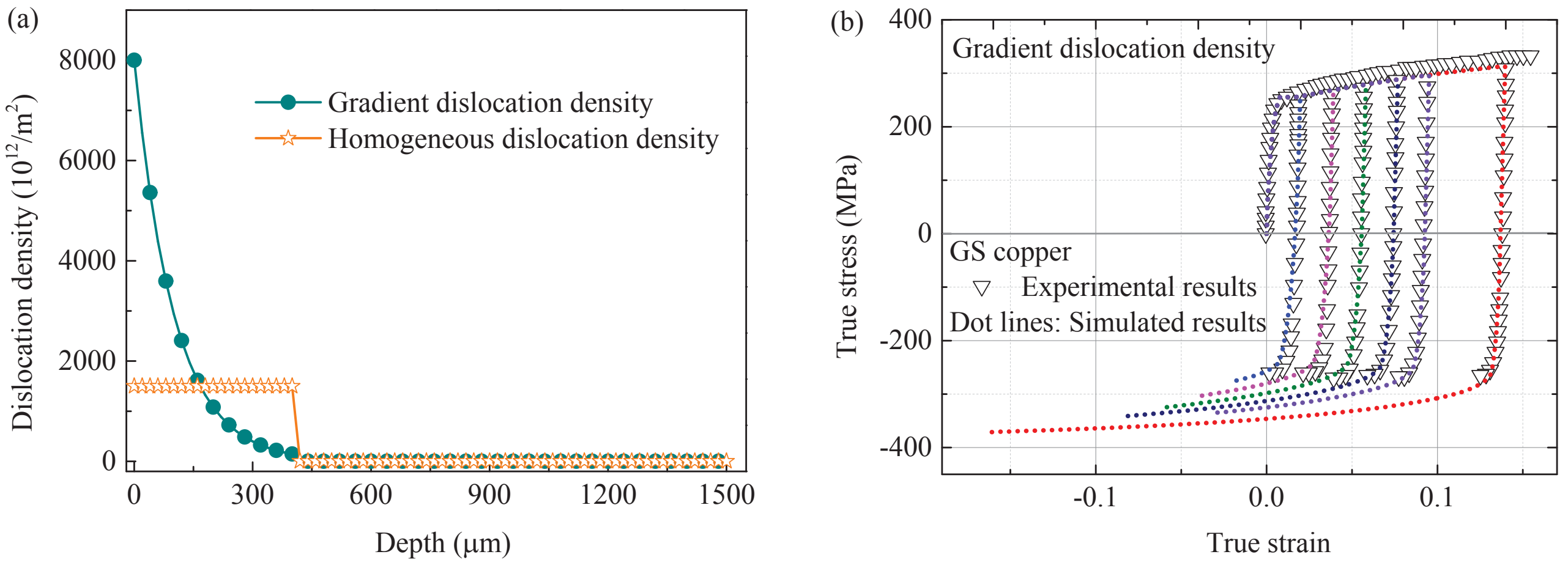

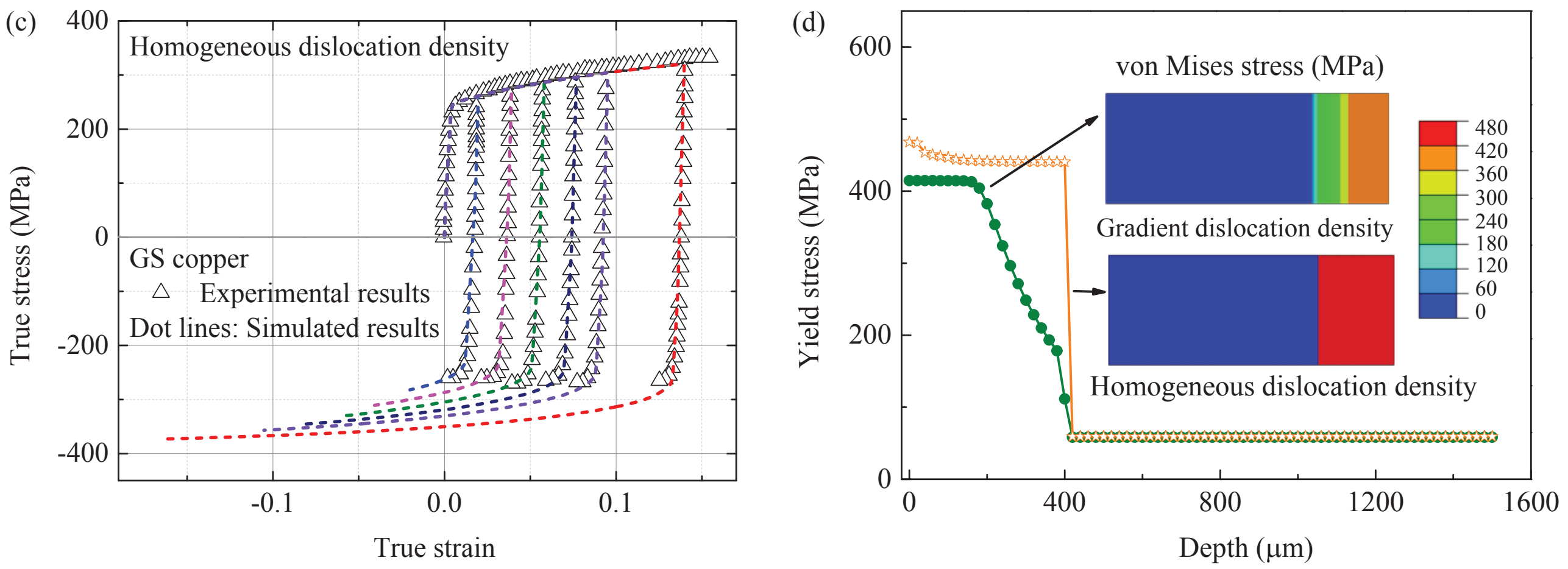

**Fig. 5.** (a) Initial dislocation density profiles adopted in the simulations. (b) Simulated tension-compression responses of GS copper with continuously graded initial dislocation density and (c) with piecewise constant initial dislocation density. The respective comparisons with experimental data are also shown. (d) The distribution of $\sigma_{0.2}$ by using continuously graded and piecewise constant initial dislocation densities.

As shown in Fig. 5(b) and 5(c), both gradient dislocation density and piecewise constant density can well predict the experimental curves for tension and compression. This result coincides with that of previous models, i.e., a homogeneous initial dislocation density can well reproduce the stress-strain curves if the dislocation density is adjusted (Li et al., 2017; Lu et al., 2019; Zhao et al., 2019a). However, problems arise when we further inspect some features characterizing the stress field, such as the hardness along the depth direction. Fig. 5(d) shows the distribution of $\sigma_{0.2}$ along the depth direction. For the case of continuously graded initial dislocation density, the yield stress decreases from the surface to the CG core in a gradual manner, which agrees with the general distribution of hardness for material processed by severe plastic deformation (Fang et al., 2014; Wu et al., 2014a; Yang et al., 2015; Yin et al., 2016). For the homogeneous case, there is a jump of yield stress at the border between the gradient layer and CG core, which is inconsistent with experimental data. This discontinuity of strength may have important repercussions since it implies that unphysical stress concentrations may emerge during loading, which in turn affect the simulated fatigue and fracture behavior. Therefore, although an initial homogeneous dislocation density in gradient layer is capable of providing a good agreement with experimental results in terms of stress-strain curves, it may compromise the ability of the models to provide a comprehensive and exact representation of the deformation behavior of GS materials which includes aspects such as damage and failure.

### 4.2.2. Kinematic hardening and Bauschinger effect

#### 4.2.2.1. Effect of back stress

The simulated tension-compression responses of CG and GS copper with and without back stress are shown in Fig. 6(a) and 6(b), respectively. For CG copper, the back stress has only little influence on the tension-compression curves, even the simulated results without back stress can also provide an acceptable prediction of the experimental results. In comparison, the modeling results without back stress significantly underrate the strain hardening of GS copper. This observation emphasizes the importance of back stresses in strain hardening of the graded material. As shown in Fig. 7(a), in CG copper, the back stress increases from 0 MPa to about 15.3 MPa when plastic strain increases from 0 to 26.2%, whereas in GS copper, the back stress increases to 29.1 MPa, indicating stronger kinematic hardening.

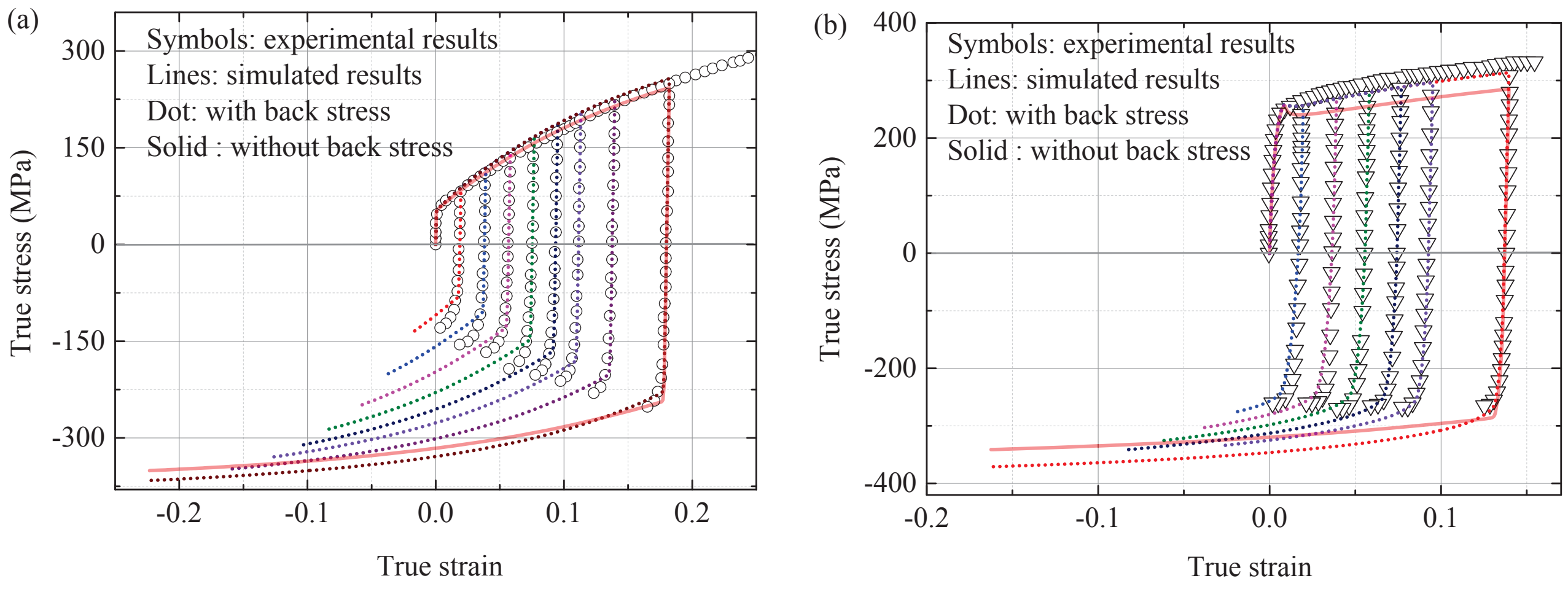

**Fig. 6.** Simulated tension-compression curves with and without back stress for (a) CG and (b) GNG copper.

To understand the mechanisms underlying the higher back stress in GS materials, Fig. 7(b) plots the evolution of back stress with plastic strain for different grain sizes. It is evident that copper with smaller grains displays larger back stress. This conclusion agrees with that obtained by Groma et al.

(Groma et al., 2003) using a continuum description of dislocations and Evers et al. (Evers et al., 2004) using size-dependent crystal plasticity: a smaller volume within which dislocation motion is confined implies higher back stress. In polycrystals, as studied here, grain boundaries act as the constraint for dislocation movement, so smaller grains exhibit higher back stress. Experimental results also demonstrated that copper with smaller grains poses a more significant Bauschninger effect (Mahato et al., 2016; Vinogradov et al., 1997). Therefore, it is concluded that the higher back stress and superior Bauschinger effect in GS materials than that in CG materials are mainly contributed by small grains in the GS layer. A number of experimental works (Park et al., 2018; Shin et al., 2019; Wang et al., 2019b; Wu et al., 2015) claimed that in heterogeneous materials with distinct differences in mechanical properties of adjacent phases (e.g., grains, layers), severe dislocation pile-ups were generated in the phase with lower yield strength and contributed a lot to kinematic (back stress) hardening as well as forest hardening, indicating an extra strengthening mechanism. However, these scenarios are of limited importance in GS materials since adjacent grains show only small size differences and thus their yield stresses are similar. So the pile-up behavior of dislocations is not significantly altered by yield stress gradient effects.

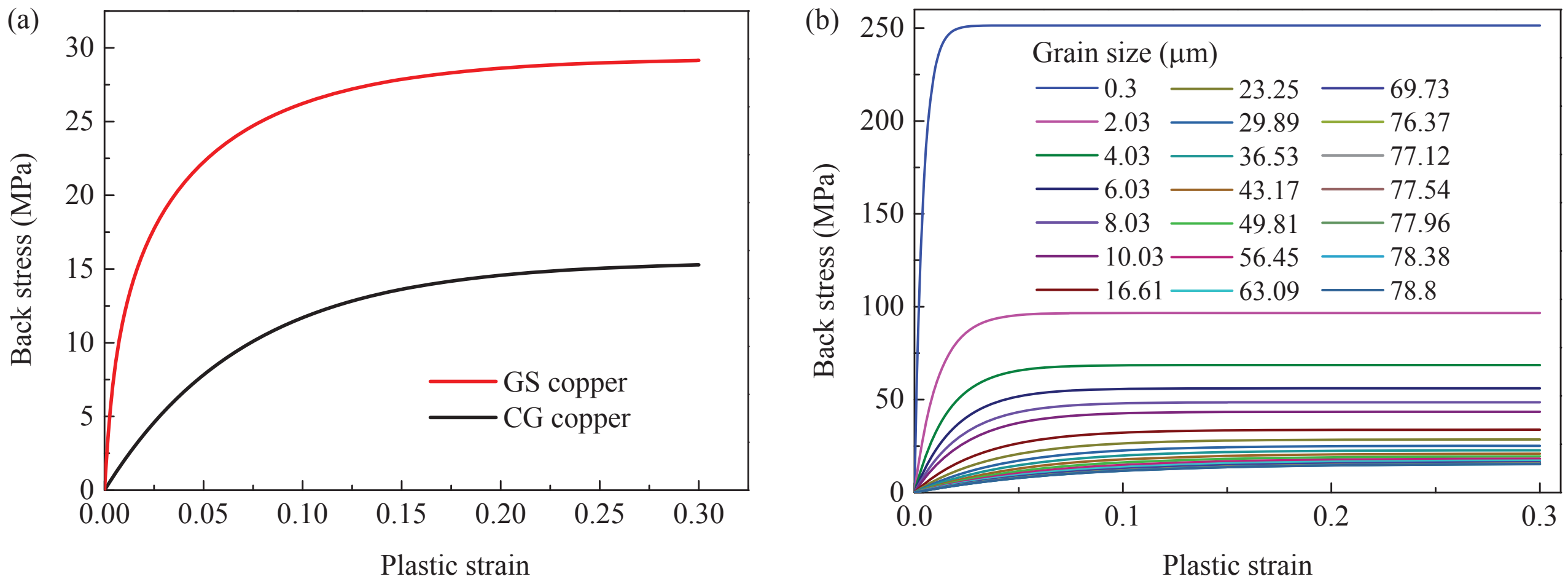

**Fig. 7.** The evolution of back stresses (a) for CG and GS copper and (b) for copper with grain sizes range from 300 nm to 78.8 μm for uniaxial loading.

#### 4.2.2.2. Effect of reversible dislocations

To investigate the reversible dislocations that may affect the non-linearity of stress-strain curves during unloading and Bauschinger effect, Fig. 8(a) compares the tension-compression behaviors of the CG, UFG and GS copper with and without pile-up GNDs density. We observe that the stress-strain curves without pile-up GNDs density show negligible differences with the curves considering pile-up GNDs density for both GS and CG copper, indicating the dislocation reversible process during reverse loading has little effects on the non-linearity of unloading and on the transient in strain hardening rate during reverse loading. For UFG copper with a grain size of 500 nm, the pile-up GNDs affect the flow stress during both tension and compression. The reason for the phenomena in Fig. 8(a) can be interpreted through the dislocation density evolutions. As shown in Fig. 8(b), in CG copper the density of pile-up GNDs is two orders of magnitude lower than that of SSDs, indicating the dislocation behavior and thus the flow stress in CG materials is dominated by SSDs, while the effects of pile-up GNDs can be neglected. When the grain size decreases to ultrafine-grained region, the density of pile-up GNDs increases to be comparable with that of SSDs, as shown in Fig. 8(c). However, since the CG core accounts for about 80% volume fraction of the whole GNG sample, the pile-up GNDs in CG core dominates the overall effect of pile-up GNDs density on the response of GS copper to a large extent. Note that in this work the concept of reversible dislocations is a little different from that reported in the literature (Castelluccio and McDowell, 2017; Kitayama et al., 2013; Rauch et al., 2007; Wen et al., 2015; Zecevic and Knezevic, 2015), in which they are considered as a part of stored dislocations during preloading and come into effect when the loading reverses or strain path changes. The concept of reversible dislocations in this work corresponds to the physical process of dislocation pile-ups near grain boundaries: during unloading and further reverse loading, the formerly formed dislocation pile-ups are released and move towards the opposite grain boundary, without any ambiguities.

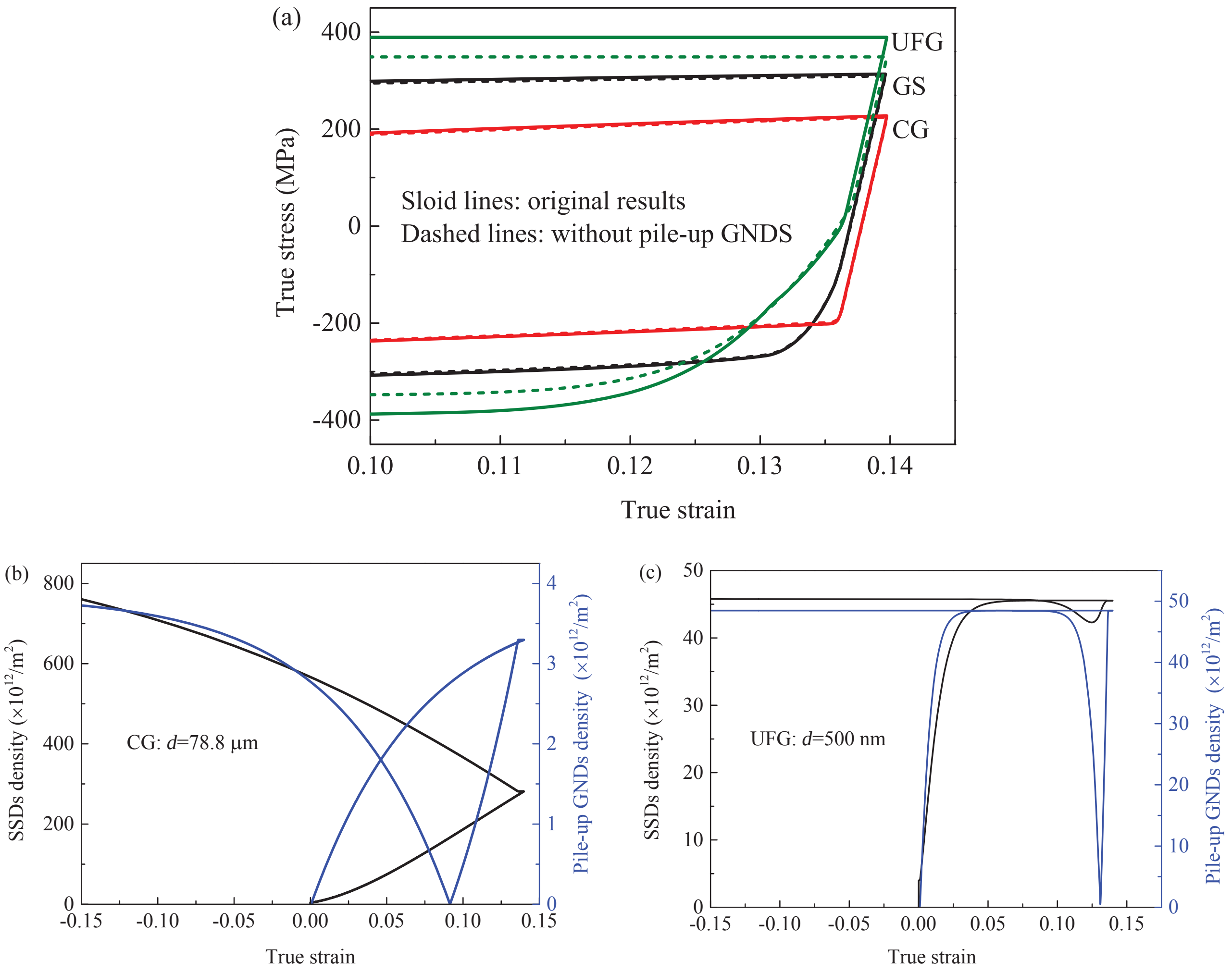

**Fig. 8.** (a) Simulated tension-compression curves of the CG and GS copper with and without pile-up GNDs density. (b) Evolution of pile-up GNDs density and SSDs density for CG and UFG copper during tension-compression.

In the pure copper studied here, grain boundary serves as the main obstacle for dislocation moving, the dislocation cell-wall structures inside grains are not considered explicitly. The release of pile-ups during reverse loading also shows an insignificant effect on the transient in strain hardening rate, which coincides with the experimental results on the cyclic deformation of polycrystalline copper (Chen and Lu, 2007; Vinogradov et al., 1997), i.e., no obvious strain hardening rate change is observed. The importance of reversible dislocation density becomes evident in alloys with the second phase inside to serve as obstacles for dislocation moving, and thus the back stress as well as the reversible dislocation density may increase (Rauch et al., 2007; Wen et al., 2015; Zecevic and Knezevic, 2015). Although the reversible dislocation density has no effect here, the model established enables a physical description of the reversible dislocation density, correlates it with the back stress and GNDs accommodating grain level plastic inhomogeneities, without inducing independent evolution laws and the related parameters.

## 5. Conclusions

A deformation-mechanism-based strain gradient plasticity model considering the plasticity heterogeneities from the grain to the sample scale is developed to describe the tension-compression response of GS materials. The dependences of the grain scale GNDs accommodating internal plasticity heterogeneities and the resulting back stress on grain size are constructed physically. It is shown that the established model integrates the grain scale GNDs, the resulting back stress, and reversible dislocation density during reverse loading into the same framework without introducing independent evolution laws and expedient parameters. A finite element implementation of the developed model successfully predicts the uniaxial tensile and tension-compression responses of GS copper as well as of CG samples. It is found that although an initial homogeneous dislocation density can also provide a good prediction in terms of the stress-strain curves, however, this assumption implies an unphysical discontinuity of the yield stress between the gradient layer and CG core. Furthermore, the back stress is demonstrated to yield a considerable contribution to strain hardening of the GS material, while back stress effects in CG copper can be neglected. The higher back stress in GS materials is mainly contributed by the fine grains in the GS layer.

## Acknowledgments

This work was supported by the National Natural Science Foundation of China (Grant Nos. 11672251 and 11872321) and the Opening fund of State Key Laboratory of Nonlinear Mechanics.

## Appendix A: numerical implementation process of the developed model

*A1. The semi-implicit stress integration scheme*

Applying the backward Euler method to the flow rule of Eqs. (4) and (12), a semi-implicit stress integration scheme is given below.

Considering the interval from step *n* to step *n*+1 in a time increment $\Delta t_{n+1} = t_{n+1} - t_n$, all information at step $n$ have been obtained, $\Delta t_{n+1}$ and $\Delta\varepsilon_{n+1}$ are given, and now we aim to obtain the stress $\boldsymbol{\sigma}_{n+1}$, the dislocation density $\rho_{\mathrm{SSDs},\,n+1}$ and $\rho_{\mathrm{GNDs},\,n+1}$ at step $n$+1. During the calculation of $\rho_{\mathrm{GNDs}}$, since the information of the neighboring integration points is required, so the explicate integration scheme is used, and implicate integration scheme is applied for calculating $\boldsymbol{\sigma}_{n+1}$ and $\rho_{\mathrm{SSDs}}$.

The stress at step $n$+1 is,

$$\boldsymbol{\sigma}_{n+1} = \boldsymbol{D} : \left(\boldsymbol{\varepsilon}_{n+1} - \boldsymbol{\varepsilon}^{\mathrm{p}}_{n+1}\right), \tag{A28}$$

where $\boldsymbol{D}$ is the elastic matrix. The deviatoric stress tensor $\boldsymbol{\sigma}^{'}_{n+1} = 2G\left(\boldsymbol{\varepsilon}^{'}_{n+1} - \boldsymbol{\varepsilon}^{\mathrm{p}}_{n+1}\right)$. The plastic strain increment is related to the effective plastic strain increment by

$$\Delta\boldsymbol{\varepsilon}^{\mathrm{p}}_{n+1} = \sqrt{\frac{3}{2}}\Delta p_{n+1}\boldsymbol{n}_{n+1}, \tag{A29}$$

where $\boldsymbol{n}$ is the flow direction tensor and

$$\boldsymbol{n}_{n+1} = \frac{\boldsymbol{\sigma}^{'}_{n+1} - \boldsymbol{\sigma}^{\mathrm{b}}_{n+1}}{\left|\boldsymbol{\sigma}^{'}_{n+1} - \boldsymbol{\sigma}^{\mathrm{b}}_{n+1}\right|}. \tag{A30}$$

Define the trial stress $\boldsymbol{\sigma}^{\mathrm{tr}}_{n+1}$ as

$$\boldsymbol{\sigma}^{'\,\mathrm{tr}}_{n+1} = 2G\left(\boldsymbol{\varepsilon}^{'\mathrm{e}}_{n+1} + \boldsymbol{\varepsilon}^{\mathrm{p}}_{n+1} - \boldsymbol{\varepsilon}^{p}_{n}\right) = \boldsymbol{\sigma}^{'}_{n+1} + 2G\Delta\boldsymbol{\varepsilon}^{\mathrm{p}}_{n+1}. \tag{A31}$$

From Eq. (22) we have

$$\boldsymbol{\sigma}^{\mathrm{b}}_{n+1} = \boldsymbol{\sigma}^{\mathrm{b}}_{n} + C\Delta\boldsymbol{\varepsilon}^{\mathrm{p}}_{n+1} - \gamma\boldsymbol{\sigma}^{\mathrm{b}}_{n+1}\Delta p_{n+1} = \left(1+\gamma\Delta p_{n+1}\right)^{-1}\left(\boldsymbol{\sigma}^{\mathrm{b}}_{n} + C\Delta\boldsymbol{\varepsilon}^{\mathrm{p}}_{n+1}\right). \tag{A32}$$

Note that for conciseness, we use the terminology of Eq. (22) in this derivation. In the implementation of the model, $C$ and $\gamma$ are replaced by the corresponding terms $8M\mu\lambda\big/\left(9\pi\left(1-\nu\right)d\right)$ and $4\mu\lambda\big/\left(3\pi\left(1-\nu\right)k_{\mathrm{HP}}\sqrt{d}\right)$ of Eq. (21).

Combining Eqs. (A31) and (A32) yields

$$\begin{aligned}\boldsymbol{\sigma}^{'}_{n+1} - \boldsymbol{\sigma}^{\mathrm{b}}_{n+1} &= \boldsymbol{\sigma}^{'\,\mathrm{tr}}_{n+1} - \left(1+\gamma\Delta p_{n+1}\right)^{-1}\left(\boldsymbol{\sigma}^{\mathrm{b}}_{n} + C\Delta\boldsymbol{\varepsilon}^{\mathrm{p}}_{n+1}\right) - 2G\Delta\boldsymbol{\varepsilon}^{\mathrm{p}}_{n+1} \\ &= \left[\boldsymbol{\sigma}^{'\,\mathrm{tr}}_{n+1} - \left(1+\gamma\Delta p_{n+1}\right)^{-1}\boldsymbol{\sigma}^{\mathrm{b}}_{n}\right] - \left[C\left(1+\gamma\Delta p_{n+1}\right)^{-1} + 2G\right]\Delta\boldsymbol{\varepsilon}^{\mathrm{p}}_{n+1}\end{aligned}. \tag{A33}$$

Since $\boldsymbol{\sigma}'_{n+1}-\boldsymbol{\sigma}^{\mathrm{b}}_{n+1}$ and $\Delta\boldsymbol{\varepsilon}^{\mathrm{p}}_{n+1}$ are coaxial, so the direction of $\left[\boldsymbol{\sigma}'^{\mathrm{tr}}_{n+1}-\left(1+\gamma\Delta p_{n+1}\right)^{-1}\boldsymbol{\sigma}^{\mathrm{b}}_{n}\right]$ should also be $\boldsymbol{n}_{n+1}$. Multiply both sides of Eq. (A33) by $\sqrt{3}\boldsymbol{n}_{n+1}/\sqrt{2}$ we have

$$\bar{\sigma}_{n+1}=\sqrt{\frac{3}{2}}\left|\boldsymbol{\sigma}'^{\mathrm{tr}}_{n+1}-\left(1+\gamma\Delta p_{n+1}\right)^{-1}\boldsymbol{\sigma}^{\mathrm{b}}_{n}\right|-\frac{3}{2}\left[C\left(1+\gamma\Delta p_{n+1}\right)^{-1}+2G\right]\Delta p_{n+1}. \tag{A34}$$

From Eq. (4) the relation between the effective plastic strain increment and effective stress can be obtained as

$$\Delta p_{n+1}=\Delta\bar{\varepsilon}_{n+1}\left(\frac{\bar{\sigma}_{n+1}}{\sigma^{\mathrm{f}}_{n+1}}\right)^{m}. \tag{A35}$$

Substituting Eq. (A34) into (A35) gives

$$\Delta p_{n+1}=\Delta\bar{\varepsilon}_{n+1}\left(\frac{\sqrt{\frac{3}{2}}\left|\boldsymbol{\sigma}'^{\mathrm{tr}}_{n+1}-\left(1+\gamma\Delta p_{n+1}\right)^{-1}\boldsymbol{\sigma}^{\mathrm{b}}_{n}\right|-\frac{3}{2}\left[C\left(1+\gamma\Delta p_{n+1}\right)^{-1}+2G\right]\Delta p_{n+1}}{\sigma_{\mathrm{f},\,n+1}}\right)^{m}. \tag{A36}$$

From the definition of flow stress of Eq. (11) we get

$$\sigma_{\mathrm{f},\,n+1}=\sigma_{0}+k_{\mathrm{HP}}d^{-1/2}+M\alpha\mu b\sqrt{\rho_{\mathrm{SSDs},\,n+1}+\rho_{\mathrm{GNDs},\,n+1}}. \tag{A37}$$

As stated before, we use an explicit integration scheme to calculate $\rho_{\mathrm{GNDs}}$, so substituting Eq. (A34) into (A35) yields

$$\Delta p_{n+1}=\Delta\bar{\varepsilon}_{n+1}\left(\frac{\sqrt{\frac{3}{2}}\left|\boldsymbol{\sigma}'^{\mathrm{tr}}_{n+1}-\left(1+\gamma\Delta p_{n+1}\right)^{-1}\boldsymbol{\sigma}^{\mathrm{b}}_{n}\right|-\frac{3}{2}\left[C\left(1+\gamma\Delta p_{n+1}\right)^{-1}+2G\right]\Delta p_{n+1}}{\sigma_{0}+k_{\mathrm{HP}}d^{-1/2}+M\alpha\mu b\sqrt{\rho_{\mathrm{SSDs},\,n}+\rho_{\mathrm{GNDs},\,n}+\Delta\rho_{\mathrm{SSDs},\,n+1}}}\right)^{m}, \tag{A38}$$

From the evolution of the SSDs density, i.e., Eq. (25), we have

$$\Delta\rho_{\mathrm{SSDs},\,n+1}=M\begin{bmatrix}\dfrac{k^{\mathrm{g}}_{\mathrm{mfp}}}{bd}+\dfrac{k^{\mathrm{dis}}_{\mathrm{mfp}}}{b}\sqrt{\rho_{\mathrm{SSDs},\,n}+\rho_{\mathrm{GNDs},\,n}+\Delta\rho_{\mathrm{SSDs},\,n+1}}\\ -k^{0}_{\mathrm{ann}}\left(\dfrac{\Delta p_{n+1}}{\dot{\varepsilon}_{\mathrm{ref}}\Delta t}\right)^{-\frac{1}{n_0}}\left(\rho_{\mathrm{SSDs},\,n}+\rho_{\mathrm{GNDs},\,n}+\Delta\rho_{\mathrm{SSDs},\,n+1}\right)\\ -\left(\dfrac{d_{\mathrm{ref}}}{d}\right)^{2}\left(\rho_{\mathrm{SSDs},\,n}+\rho_{\mathrm{GNDs},\,n}+\Delta\rho_{\mathrm{SSDs},\,n+1}\right)\end{bmatrix}\Delta p_{n+1}. \tag{A39}$$

Combining Eqs. (A38) and (A39) we get a system of nonlinear equations with respect to $\Delta p_{n+1}$ and $\Delta\rho_{\mathrm{SSDs},\,n+1}$. Let

$$f\left(\Delta p_{n+1},\Delta\rho_{\mathrm{SSDs},\,n+1}\right)=\Delta p_{n+1}-\Delta\bar{\varepsilon}_{n+1}\left(\frac{\sqrt{\frac{3}{2}}\left|\boldsymbol{\sigma}'^{\mathrm{tr}}_{n+1}-\left(1+\gamma\Delta p_{n+1}\right)^{-1}\boldsymbol{\sigma}^{\mathrm{b}}_{n}\right|-\frac{3}{2}\left[C\left(1+\gamma\Delta p_{n+1}\right)^{-1}+2G\right]\Delta p_{n+1}}{\sigma_0+k_{\mathrm{HP}}d^{-1/2}+M\alpha\mu b\sqrt{\rho_{\mathrm{SSDs},\,n}+\rho_{\mathrm{GNDs},\,n}+\Delta\rho_{\mathrm{SSDs},\,n+1}}}\right)^{m}$$

$$g\left(\Delta p_{n+1},\Delta\rho_{\mathrm{SSDs},\,n+1}\right)=\Delta\rho_{\mathrm{SSDs},\,n+1}-M\begin{bmatrix}\frac{k^{\mathrm{g}}_{\mathrm{mfp}}}{bd}+\frac{k^{\mathrm{dis}}_{\mathrm{mfp}}}{b}\sqrt{\rho_{\mathrm{SSDs},\,n}+\rho_{\mathrm{GNDs},\,n}+\Delta\rho_{\mathrm{SSDs},\,n+1}}\\ -k^{0}_{\mathrm{ann}}\left(\frac{\Delta p_{n+1}}{\dot{\varepsilon}_{\mathrm{ref}}\Delta t}\right)^{-\frac{1}{n_0}}\left(\rho_{\mathrm{SSDs},\,n}+\rho_{\mathrm{GNDs},\,n}+\Delta\rho_{\mathrm{SSDs},\,n+1}\right)\\ -\left(\frac{d_{\mathrm{ref}}}{d}\right)^{2}\left(\rho_{\mathrm{SSDs},\,n}+\rho_{\mathrm{GNDs},\,n}+\Delta\rho_{\mathrm{SSDs},\,n+1}\right)\end{bmatrix}\Delta p_{n+1} \quad \text{(A40)}$$

Then $\Delta p_{n+1}$ and $\Delta\rho_{\mathrm{SSDs},\,n+1}$ can be obtained by solving this system of nonlinear equations using the Newton-Raphson iterative method.

### *A3. Consistent tangent modulus*

For the finite element implementation of the constitutive model, the consistent tangent modulus used can be expressed as

$$\boldsymbol{C}=\frac{\partial\Delta\boldsymbol{\sigma}_{n+1}}{\partial\Delta\boldsymbol{\varepsilon}_{n+1}}. \quad \text{(A41)}$$

From Eq. (A28) we have

$$\mathrm{d}\Delta\boldsymbol{\sigma}_{n+1}=\boldsymbol{D}:\left(\mathrm{d}\Delta\boldsymbol{\varepsilon}_{n+1}-\mathrm{d}\Delta\boldsymbol{\varepsilon}^{\mathrm{p}}_{n+1}\right). \quad \text{(A42)}$$

To obtain the consistent tangent modulus ***C***, a relation between $\mathrm{d}\Delta\boldsymbol{\varepsilon}_{n+1}$ and $\mathrm{d}\Delta\boldsymbol{\varepsilon}^{p}_{n+1}$ has to be derived. From Eq. (A29) we have

$$\mathrm{d}\Delta\boldsymbol{\varepsilon}^{\mathrm{p}}_{n+1}=\sqrt{\frac{3}{2}}\left(\mathrm{d}\Delta p_{n+1}\boldsymbol{n}_{n+1}+\Delta p_{n+1}\mathrm{d}\boldsymbol{n}_{n+1}\right). \quad \text{(A43)}$$

So $\mathrm{d}\Delta p_{n+1}$ and $\mathrm{d}\boldsymbol{n}_{n+1}$ need to be found. From Eq. (A35),

$$\mathrm{d}\Delta p_{n+1}=\mathrm{d}\Delta\bar{\varepsilon}_{n+1}\left(\frac{\bar{\sigma}_{n+1}}{\sigma_{\mathrm{f},\,n+1}}\right)^{m}+\frac{m\Delta\bar{\varepsilon}_{n+1}}{\sigma_{\mathrm{f},\,n+1}}\left(\frac{\bar{\sigma}_{n+1}}{\sigma_{\mathrm{f},\,n+1}}\right)^{m-1}\mathrm{d}\bar{\sigma}_{n+1}-\frac{m\Delta\bar{\varepsilon}_{n+1}\bar{\sigma}_{n+1}}{\left(\sigma_{\mathrm{f},\,n+1}\right)^{2}}\left(\frac{\bar{\sigma}_{n+1}}{\sigma_{\mathrm{f},\,n+1}}\right)^{m-1}\mathrm{d}\sigma_{\mathrm{f},\,n+1}, \quad \text{(A44)}$$

and from Eq. (A30),

$$\mathrm{d}\boldsymbol{n}_{n+1}=\left[\left|\boldsymbol{\sigma}'_{n+1}-\boldsymbol{\sigma}^{\mathrm{b}}_{n+1}\right|^{-1}\left(\boldsymbol{I}-\boldsymbol{n}_{n+1}\otimes\boldsymbol{n}_{n+1}\right)\right]:\left(\mathrm{d}\Delta\boldsymbol{\sigma}'_{n+1}-\mathrm{d}\Delta\boldsymbol{\sigma}^{\mathrm{b}}_{n+1}\right), \tag{A45}$$

with $\boldsymbol{I}$ bening the fourth-order symmetric unit tensor.

Since $\Delta\bar{\varepsilon}_{n+1}=\sqrt{2/\left(3\Delta\boldsymbol{\varepsilon}_{n+1}:\Delta\boldsymbol{\varepsilon}_{n+1}\right)}$ , so

$$\mathrm{d}\Delta\bar{\varepsilon}_{n+1}=\frac{2}{3}\frac{\Delta\boldsymbol{\varepsilon}_{n+1}}{\Delta\bar{\varepsilon}_{n+1}}:\mathrm{d}\Delta\boldsymbol{\varepsilon}_{n+1}, \tag{A46}$$

and $\mathrm{d}\bar{\sigma}_{n+1}$ can also be obtained as

$$\mathrm{d}\bar{\sigma}_{n+1}=\frac{3}{2}\frac{\boldsymbol{\sigma}'_{n+1}-\boldsymbol{\sigma}^{\mathrm{b}}_{n+1}}{\bar{\sigma}_{n+1}}:\left(\mathrm{d}\boldsymbol{\sigma}'_{n+1}-\mathrm{d}\boldsymbol{\sigma}^{\mathrm{b}}_{n+1}\right)=\frac{3}{2}\frac{\boldsymbol{\sigma}'_{n+1}-\boldsymbol{\sigma}^{\mathrm{b}}_{n+1}}{\bar{\sigma}_{n+1}}:\left(\mathrm{d}\Delta\boldsymbol{\sigma}'_{n+1}-\mathrm{d}\Delta\boldsymbol{\sigma}^{\mathrm{b}}_{n+1}\right). \tag{A47}$$

Form Eq. (A37),

$$\mathrm{d}\sigma_{\mathrm{f},\,n+1}=\frac{M\alpha\mu b}{2\sqrt{\rho_{n+1}}}\mathrm{d}\Delta\rho_{\mathrm{SSDs},\,n+1}. \tag{A48}$$

Substituting Eq. (A48) into (A44) yields

$$\begin{aligned}\mathrm{d}\Delta p_{n+1}=&\,\mathrm{d}\Delta\bar{\varepsilon}_{n+1}\left(\frac{\bar{\sigma}_{n+1}}{\sigma_{\mathrm{f},\,n+1}}\right)^{m}+\frac{m\Delta\bar{\varepsilon}_{n+1}}{\sigma_{\mathrm{f},\,n+1}}\left(\frac{\bar{\sigma}_{n+1}}{\sigma_{\mathrm{f},\,n+1}}\right)^{m-1}\mathrm{d}\bar{\sigma}_{n+1}\\&-\frac{m\Delta\bar{\varepsilon}_{n+1}\bar{\sigma}_{n+1}}{\left(\sigma_{\mathrm{f},\,n+1}\right)^{2}}\left(\frac{\bar{\sigma}_{n+1}}{\sigma_{\mathrm{f},\,n+1}}\right)^{m-1}\frac{M\alpha\mu b}{2\sqrt{\rho_{n+1}}}\mathrm{d}\Delta\rho_{\mathrm{SSDs},\,n+1}\end{aligned}. \tag{A49}$$

From Eq. (A39),

$$\begin{aligned}\mathrm{d}\Delta\rho_{\mathrm{SSDs},\,n+1}=&\,M\left[\frac{k^{\mathrm{g}}_{\mathrm{mfp}}}{bd}+\frac{k^{\mathrm{dis}}_{\mathrm{mfp}}}{b}\sqrt{\rho_{n+1}}-k^{0}_{\mathrm{ann}}\left(\frac{\Delta p_{n+1}}{\dot{\varepsilon}_{\mathrm{ref}}\Delta t}\right)^{-\frac{1}{n_0}}\rho_{n+1}-\left(\frac{d_{\mathrm{ref}}}{d}\right)^{2}\rho_{n+1}\right]\mathrm{d}\Delta p_{n+1}\\&+M\Delta p_{n+1}\left[\begin{aligned}&\frac{k^{\mathrm{dis}}_{\mathrm{mfp}}}{2b\sqrt{\rho_{n+1}}}\mathrm{d}\Delta\rho_{\mathrm{SSDs},\,n+1}+\frac{k^{0}_{\mathrm{ann}}}{n_0\dot{\varepsilon}_{\mathrm{ref}}\Delta t}\left(\frac{\Delta p_{n+1}}{\dot{\varepsilon}_{\mathrm{ref}}\Delta t}\right)^{-\frac{1+n_0}{n_0}}\rho_{n+1}\mathrm{d}\Delta p_{n+1}\\&-k^{0}_{\mathrm{ann}}\left(\frac{\Delta p_{n+1}}{\dot{\varepsilon}_{\mathrm{ref}}\Delta t}\right)^{-\frac{1}{n_0}}\mathrm{d}\Delta\rho_{\mathrm{SSDs},\,n+1}-\left(\frac{d_{\mathrm{ref}}}{d}\right)^{2}\mathrm{d}\Delta\rho_{\mathrm{SSDs},\,n+1}\end{aligned}\right]\end{aligned}. \tag{A50}$$

So,

$$\mathrm{d}\Delta\rho_{\mathrm{SSDs},\,n+1}=\frac{\left\{\begin{array}{l}M\left[\dfrac{k_{\mathrm{mfp}}^{\mathrm{g}}}{bd}+\dfrac{k_{\mathrm{mfp}}^{\mathrm{dis}}}{b}\sqrt{\rho_{n+1}}-k_{\mathrm{ann}}^{0}\left(\dfrac{\Delta p_{n+1}}{\dot{\varepsilon}_{\mathrm{ref}}\Delta t}\right)^{-\frac{1}{n_0}}\rho_{n+1}-\left(\dfrac{d_{\mathrm{ref}}}{d}\right)^{2}\rho_{n+1}\right]\\+\dfrac{M\Delta p_{n+1}k_{\mathrm{ann}}^{0}}{n_0\dot{\varepsilon}_{\mathrm{ref}}\Delta t}\left(\dfrac{\Delta p_{n+1}}{\dot{\varepsilon}_{\mathrm{ref}}\Delta t}\right)^{-\frac{1+n_0}{n_0}}\rho_{n+1}\end{array}\right\}}{1-\dfrac{M\Delta p_{n+1}k_{\mathrm{mfp}}^{\mathrm{dis}}}{2b\sqrt{\rho_{n+1}}}+M\Delta p_{n+1}\left(\dfrac{d_{\mathrm{ref}}}{d}\right)^{2}+M\Delta p_{n+1}k_{\mathrm{ann}}^{0}\left(\dfrac{\Delta p_{n+1}}{\dot{\varepsilon}_{\mathrm{ref}}\Delta t}\right)^{-\frac{1}{n_0}}}\mathrm{d}\Delta p_{n+1}. \quad \text{(A51)}$$

For conciseness let

$$\mathrm{A}=\frac{\left\{\begin{array}{l}M\left[\dfrac{k_{\mathrm{mfp}}^{\mathrm{g}}}{bd}+\dfrac{k_{\mathrm{mfp}}^{\mathrm{dis}}}{b}\sqrt{\rho_{n+1}}-k_{\mathrm{ann}}^{0}\left(\dfrac{\Delta p_{n+1}}{\dot{\varepsilon}_{\mathrm{ref}}\Delta t}\right)^{-\frac{1}{n_0}}\rho_{n+1}-\left(\dfrac{d_{\mathrm{ref}}}{d}\right)^{2}\rho_{n+1}\right]\\+\dfrac{M\Delta p_{n+1}k_{\mathrm{ann}}^{0}}{n_0\dot{\varepsilon}_{\mathrm{ref}}\Delta t}\left(\dfrac{\Delta p_{n+1}}{\dot{\varepsilon}_{\mathrm{ref}}\Delta t}\right)^{-\frac{1+n_0}{n_0}}\rho_{n+1}\end{array}\right\}}{1-\dfrac{M\Delta p_{n+1}k_{\mathrm{mfp}}^{\mathrm{dis}}}{2b\sqrt{\rho_{n+1}}}+M\Delta p_{n+1}\left(\dfrac{d_{\mathrm{ref}}}{d}\right)^{2}+M\Delta p_{n+1}k_{\mathrm{ann}}^{0}\left(\dfrac{\Delta p_{n+1}}{\dot{\varepsilon}_{\mathrm{ref}}\Delta t}\right)^{-\frac{1}{n_0}}}, \quad \text{(A52)}$$

then

$$\mathrm{d}\Delta\rho_{\mathrm{SSDs},\,n+1}=A\mathrm{d}\Delta p_{n+1}. \quad \text{(A53)}$$

Substituting Eqs. (A46), (A47), (A48) and (A53) into (A44), then let

$$\begin{array}{l}\boldsymbol{J}_1=\dfrac{2}{3\Delta\bar{\varepsilon}_{n+1}}\left(\dfrac{\bar{\sigma}_{n+1}}{\sigma_{\mathrm{f},\,n+1}}\right)^{m}\Delta\boldsymbol{\varepsilon}_{n+1},\ \boldsymbol{J}_2=\dfrac{3m\Delta\bar{\varepsilon}_{n+1}}{2\sigma_{\mathrm{f},\,n+1}\bar{\sigma}_{n+1}}\left(\dfrac{\bar{\sigma}_{n+1}}{\sigma_{\mathrm{f},\,n+1}}\right)^{m-1}\left(\mathrm{d}\boldsymbol{\sigma}'_{n+1}-\mathrm{d}\boldsymbol{\sigma}^{\mathrm{b}}_{n+1}\right),\\ B=\dfrac{m\Delta\bar{\varepsilon}_{n+1}\bar{\sigma}_{n+1}}{\left(\sigma_{\mathrm{f},\,n+1}\right)^{2}}\left(\dfrac{\bar{\sigma}_{n+1}}{\sigma_{\mathrm{f},\,n+1}}\right)^{m-1}\dfrac{M\alpha\mu b}{2\sqrt{\rho_{n+1}}}A\end{array}, \quad \text{(A54)}$$

we get

$$\mathrm{d}\Delta p_{n+1}=\frac{1}{1+B}\left[\boldsymbol{J}_1:\mathrm{d}\Delta\boldsymbol{\varepsilon}_{n+1}+\boldsymbol{J}_2:\left(\mathrm{d}\Delta\boldsymbol{\sigma}'_{n+1}-\mathrm{d}\Delta\boldsymbol{\sigma}^{\mathrm{b}}_{n+1}\right)\right]. \quad \text{(A55)}$$

Further, defining $\boldsymbol{L}_1=\boldsymbol{J}_1/(1+B),\ \boldsymbol{L}_2=\boldsymbol{J}_2/(1+B)$ yields

$$\mathrm{d}\Delta p_{n+1}=\boldsymbol{L}_1:\mathrm{d}\Delta\boldsymbol{\varepsilon}_{n+1}+\boldsymbol{L}_2:\left(\mathrm{d}\Delta\boldsymbol{\sigma}'_{n+1}-\mathrm{d}\Delta\boldsymbol{\sigma}^{\mathrm{b}}_{n+1}\right). \quad \text{(A56)}$$

From the definition of $\boldsymbol{\sigma}'_{n+1}=2G\left(\boldsymbol{\varepsilon}'_{n+1}-\boldsymbol{\varepsilon}^{\mathrm{p}}_{n+1}\right)$, it gives

$$\mathrm{d}\Delta\boldsymbol{\sigma}'_{n+1}=2G\boldsymbol{I}_d:\mathrm{d}\Delta\boldsymbol{\varepsilon}_{n+1}-2G\mathrm{d}\Delta\boldsymbol{\varepsilon}^{\mathrm{p}}_{n+1}, \quad \text{(A57)}$$

From Eq. (A32),

$$\mathrm{d}\Delta\boldsymbol{\sigma}_{n+1}^{\mathrm{b}} = \frac{1}{1+\gamma\Delta p_{n+1}}\left(C\mathrm{d}\Delta\boldsymbol{\varepsilon}_{n+1}^{\mathrm{p}} - \gamma\boldsymbol{\sigma}_{n+1}^{\mathrm{b}}\mathrm{d}\Delta p_{n+1}\right). \tag{A58}$$

Combining Eqs. (A56), (A57), (A58) and (A42), it provides

$$\mathrm{d}\Delta p_{n+1} = \frac{1}{1-\dfrac{\gamma\boldsymbol{L}_2:\boldsymbol{\sigma}_{n+1}^{\mathrm{b}}}{1+\gamma\Delta p_{n+1}}}\left[\begin{array}{l}\left(\boldsymbol{L}_1 + 2G\boldsymbol{I}_d:\boldsymbol{L}_2 - \left(2G+\dfrac{C}{1+\gamma\Delta p_{n+1}}\right)\boldsymbol{L}_2\right):\mathrm{d}\Delta\boldsymbol{\varepsilon}_{n+1} \\ +\left(2G+\dfrac{C}{1+\gamma\Delta p_{n+1}}\right)\boldsymbol{D}^{-1}:\boldsymbol{L}_2:\mathrm{d}\Delta\boldsymbol{\sigma}_{n+1}\end{array}\right]. \tag{A59}$$

Let,

$$\begin{array}{l}\boldsymbol{M}_1 = \dfrac{1}{1-\dfrac{\gamma\boldsymbol{L}_2:\boldsymbol{\sigma}_{n+1}^{\mathrm{b}}}{1+\gamma\Delta p_{n+1}}}\left[\boldsymbol{L}_1 + 2G\boldsymbol{I}_d:\boldsymbol{L}_2 - \left(2G+\dfrac{C}{1+\gamma\Delta p_{n+1}}\right)\boldsymbol{L}_2\right], \\ \boldsymbol{M}_2 = \dfrac{1}{1-\dfrac{\gamma\boldsymbol{L}_2:\boldsymbol{\sigma}_{n+1}^{\mathrm{b}}}{1+\gamma\Delta p_{n+1}}}\left(2G+\dfrac{C}{1+\gamma\Delta p_{n+1}}\right)\boldsymbol{D}^{-1}:\boldsymbol{L}_2\end{array}, \tag{A60}$$

then,

$$\mathrm{d}\Delta p_{n+1} = \boldsymbol{M}_1:\mathrm{d}\Delta\boldsymbol{\varepsilon}_{n+1} + \boldsymbol{M}_2:\mathrm{d}\Delta\boldsymbol{\sigma}_{n+1}. \tag{A61}$$

Substituting Eqs. (A57) and (A58) into (A45), and let $\boldsymbol{H} = \left|\boldsymbol{\sigma}_{n+1}^{'} - \boldsymbol{\sigma}_{n+1}^{\mathrm{b}}\right|^{-1}\left(\boldsymbol{I} - \boldsymbol{n}_{n+1}\otimes\boldsymbol{n}_{n+1}\right)$, $E = C/\left(1+\gamma\Delta p_{n+1}\right)$, $F = \gamma/\left(1+\gamma\Delta p_{n+1}\right)$ yield

$$\mathrm{d}\boldsymbol{n}_{n+1} = \left(\boldsymbol{H}:2G\boldsymbol{I}_d\right):\mathrm{d}\Delta\boldsymbol{\varepsilon}_{n+1} - \left(2G+E\right)\boldsymbol{H}:\mathrm{d}\Delta\boldsymbol{\varepsilon}_{n+1}^{\mathrm{p}} + F\boldsymbol{H}:\boldsymbol{\sigma}_{n+1}^{\mathrm{b}}\mathrm{d}\Delta p_{n+1}. \tag{A62}$$

Substituting the expression of $\mathrm{d}\Delta p_{n+1}$, i.e., Eq.(A61), into Eq. (A62) gives

$$\begin{array}{l}\mathrm{d}\boldsymbol{n}_{n+1} = \left(\boldsymbol{H}:2G\boldsymbol{I}_d\right):\mathrm{d}\Delta\boldsymbol{\varepsilon}_{n+1} - \left(2G+E\right)\boldsymbol{H}:\mathrm{d}\Delta\boldsymbol{\varepsilon}_{n+1}^{\mathrm{p}} \\ \qquad + F\boldsymbol{H}:\boldsymbol{\sigma}_{n+1}^{\mathrm{b}}\otimes\boldsymbol{M}_1:\mathrm{d}\Delta\boldsymbol{\varepsilon}_{n+1} + F\boldsymbol{H}:\boldsymbol{\sigma}_{n+1}^{\mathrm{b}}\otimes\boldsymbol{M}_2:\mathrm{d}\Delta\boldsymbol{\sigma}_{n+1}\end{array}. \tag{A63}$$

Combing Eqs. (A61), (A63) and (A43) yields

$$\mathrm{d}\Delta\boldsymbol{\varepsilon}_{n+1}^{p} = \sqrt{\frac{3}{2}}\left\{\begin{array}{l}\boldsymbol{n}_{n+1}\otimes\boldsymbol{M}_1:\mathrm{d}\Delta\boldsymbol{\varepsilon}_{n+1} + \boldsymbol{n}_{n+1}\otimes\boldsymbol{M}_2:\mathrm{d}\Delta\boldsymbol{\sigma}_{n+1} \\ +\Delta p_{n+1}\left[\begin{array}{l}\left(\boldsymbol{H}:2G\boldsymbol{I}_d\right):\mathrm{d}\Delta\boldsymbol{\varepsilon}_{n+1} - \left(2G+E\right)\boldsymbol{H}:\mathrm{d}\Delta\boldsymbol{\varepsilon}_{n+1}^{\mathrm{p}} \\ +F\boldsymbol{H}:\boldsymbol{\sigma}_{n+1}^{\mathrm{b}}\otimes\boldsymbol{M}_1:\mathrm{d}\Delta\boldsymbol{\varepsilon}_{n+1} + F\boldsymbol{H}:\boldsymbol{\sigma}_{n+1}^{\mathrm{b}}\otimes\boldsymbol{M}_2:\mathrm{d}\Delta\boldsymbol{\sigma}_{n+1}\end{array}\right]\end{array}\right\}. \tag{A64}$$

Further considering Eq. (A42), we have

$$\mathrm{d}\Delta\boldsymbol{\varepsilon}_{n+1} - \boldsymbol{D}^{-1}:\mathrm{d}\Delta\boldsymbol{\sigma}_{n+1} = \sqrt{\frac{3}{2}}\left\{\begin{array}{l} \boldsymbol{n}_{n+1}\otimes\boldsymbol{M}_1:\mathrm{d}\Delta\boldsymbol{\varepsilon}_{n+1} + \boldsymbol{n}_{n+1}\otimes\boldsymbol{M}_2:\mathrm{d}\Delta\boldsymbol{\sigma}_{n+1} \\ +\Delta p_{n+1}\left[\begin{array}{l} \left(\boldsymbol{H}:2G\boldsymbol{I}_d\right):\mathrm{d}\Delta\boldsymbol{\varepsilon}_{n+1} - \left(2G+E\right)\boldsymbol{H}:\mathrm{d}\Delta\boldsymbol{\varepsilon}^{\mathrm{p}}_{n+1} \\ +F\boldsymbol{H}:\boldsymbol{\sigma}^{\mathrm{b}}_{n+1}\otimes\boldsymbol{M}_1:\mathrm{d}\Delta\boldsymbol{\varepsilon}_{n+1} + F\boldsymbol{H}:\boldsymbol{\sigma}^{\mathrm{b}}_{n+1}\otimes\boldsymbol{M}_2:\mathrm{d}\Delta\boldsymbol{\sigma}_{n+1} \end{array}\right] \end{array}\right\}. \quad \text{(A65)}$$

Simplifying Eq. (A65) we finally obtain the consistent tangent modulus as

$$\frac{\partial\Delta\boldsymbol{\sigma}_{n+1}}{\partial\Delta\boldsymbol{\varepsilon}_{n+1}} = \left[\begin{array}{l} \boldsymbol{D}^{-1} + \sqrt{\frac{3}{2}}\boldsymbol{n}_{n+1}\otimes\boldsymbol{M}_2 + \sqrt{\frac{3}{2}}\Delta p_{n+1}\left(2G+E\right)\boldsymbol{H}:\boldsymbol{D}^{-1} \\ +\sqrt{\frac{3}{2}}\Delta p_{n+1}F\boldsymbol{H}:\boldsymbol{\sigma}^{\mathrm{b}}_{n+1}\otimes\boldsymbol{M}_2 \end{array}\right]^{-1} : \left[\begin{array}{l} \boldsymbol{I} - \sqrt{\frac{3}{2}}\boldsymbol{n}_{n+1}\otimes\boldsymbol{M}_1 - \sqrt{\frac{3}{2}}\Delta p_{n+1}\left(\boldsymbol{H}:2G\boldsymbol{I}_d\right) + \\ \sqrt{\frac{3}{2}}\Delta p_{n+1}\left(2G+E\right)\boldsymbol{H} - \sqrt{\frac{3}{2}}\Delta p_{n+1}F\boldsymbol{H}:\boldsymbol{\sigma}^{\mathrm{b}}_{n+1}\otimes\boldsymbol{M}_1 \end{array}\right]. \quad (66)$$

## Appendix B: determination of parameters in the developed model

Since the flow stresses in Fig. 4(a), Fig. 4(b) Fig. 5(b) and Fig. 5(c) all include back stress, the parameters related to back stress are discussed first. As indicated by Eqs. (14), (15), (17) and (20), back stress can be related to the parameters $k_{\mathrm{HP}}$ and $\lambda$, where $k_{\mathrm{HP}}$ relates to the level of back stress while $\lambda$ to its evolution rate. Here $k_{\mathrm{HP}}$ is obtained by evaluating the back stress level for CG copper with a grain size of 78.8 μm. Once $k_{\mathrm{HP}}$ is determined, the back stress levels of both CG and GS coppers are decided. Then $\lambda$ is determined by evaluating the back stress evolution rate for both CG and GS copper. Note that $\lambda$ is the distance between slip lines and should have a physical value, e.g., the value can not be so large that it even surpasses the size of a CG and also can not be too small that it falls below the magnitude of Burgers vector. Sinclair et al. (Sinclair et al., 2006) obtained this value as 413 nm, Bouaziz (Bouaziz et al., 2008) et al. obtained it as 1266$b$ ($\approx$ 316.5 nm) and Li et al. (Li and Soh, 2012) set it as 200 nm. In this work, we also obtain this value as 200 nm, which is the same as that adopted by Li et al. (Li and Soh, 2012).

Some parameters related to the evolution of SSDs' density shown in Table 1 can be extracted from the literature. Here we only discuss those that need to be determined in this work, i.e., the parameters controlling the evolution of SSDs' density $k^{\mathrm{g}}_{\mathrm{mfp}}$, $k^{\mathrm{dis}}_{\mathrm{mfp}}$, $k_{\mathrm{ann}}$ and $d_{\mathrm{ref}}$. Although four parameters need to be determined, this is not difficult since different parameters ar relevant in different grain size regimes. For example, when the grain size is large enough, the influence of the first and fourth terms related to grain size in Eq. (25) on the evolution of SSD density can be neglected. So the evolution of $\rho_{\mathrm{SSDs}}$ is merely controlled by $k_{\mathrm{mfp}}^{\mathrm{dis}}$ and $k_{\mathrm{ann}}$. Correspondingly, when the grain size is small enough, the influence of the second and third terms can be neglected and the evolution of $\rho_{\mathrm{SSDs}}$ is mainly controlled by $k_{\mathrm{mfp}}^{\mathrm{g}}$ and $d_{\mathrm{ref}}$. In this work, $k_{\mathrm{mfp}}^{\mathrm{dis}}$ and $k_{\mathrm{ann}}$ are determined by simulating the stress-strain curve of CG copper with a grain size of 78.8 μm in Fig. 4(a); $k_{\mathrm{mfp}}^{\mathrm{g}}$ and $d_{\mathrm{ref}}$ are obtained by simulating the copper with a grain size of 500 nm. All the four parameters are adjusted slightly for making good fits to the three stress-strain curves of 78.8 μm, 25 μm, and 500 nm simultaneously in Fig. 4(a).